\documentclass[%
aip,
amsmath,amssymb,
reprint,%
]{revtex4-1}

\usepackage{graphicx}
\usepackage{dcolumn}
\usepackage{bm}

\usepackage[utf8]{inputenc}
\usepackage[T1]{fontenc}
\usepackage{mathptmx}

\usepackage[colorlinks=false,linkcolor=blue]{hyperref}
\usepackage{xcolor, soul} 
\sethlcolor{green} 

\begin{document}

	\title{Route to logical strange nonchaotic attractors with single periodic force and noise}	

	\author{M. Sathish Aravindh} 
	\email{sathisharavindhm@gmail.com}
	\affiliation{PG \& Research Department of Physics, Nehru Memorial College (Autonomous), Affiliated to Bharathidasan University, Puthanampatti, Tiruchirappalli - 621 007, India.}
	
	\affiliation{Department of Nonlinear Dynamics, School of Physics, Bharathidasan University, Tiruchirappalli - 620 024, India.}
	\author{A. Venkatesan}
	\email{av.phys@gmail.com}
	\affiliation{PG \& Research Department of Physics, Nehru Memorial College (Autonomous), Affiliated to Bharathidasan University, Puthanampatti, Tiruchirappalli - 621 007, India.}

	\author{M. Lakshmanan}
	\email{lakshman.cnld@gmail.com}
	\affiliation{Department of Nonlinear Dynamics, School of Physics, Bharathidasan University, Tiruchirappalli - 620 024, India.}
	
	\begin{abstract}
		Strange nonchaotic attractors (SNAs) have been identified and studied in the  literature exclusively in quasiperiodically driven nonlinear dynamical systems. It is an interesting question to ask whether they can be identified with other types of forcings as well, which still remains as an open problem. Here, we show that robust SNAs can be created by a small amount of noise in periodically driven nonlinear dynamical systems by a single force. The robustness of these attractors is tested by perturbing the system with logical signals leading to the emulation of different logical elements in the SNA regions.
	\end{abstract}
	\maketitle

	\begin{quotation}
	The question whether strange nonchaotic attractors (SNAs) can occur typically in nonlinear dynamical systems other than the quasiperiodically forced ones still remains as an open problem. In this paper, we show that SNAs can be generated by a small amount of noise in a periodically driven Duffing oscillator with a single force. Robustness of the resulting SNA phenomenon can be verified by perturbing the system with logical signals. It is interesting to note that robust SNAs with logical behavior (logical SNA) and without logical behavior (standard SNA) are observed with different input streams. The logical behavior in the SNA regime is robust in the presence of experimental noise too. Thus the present study paves the way to construct alternative computing in reliable and reconfigurable computer architecture.
	\end{quotation}

	\section{Introduction} 
	
Strange nonchaotic attractor(SNA) is an attractor which has a complicated geometry but its maximal Lyapunov exponent is not positive and therefore it will not exhibit sensitive dependence on initial conditions. Such attractors were first observed by Grebogi \cite{grebogi1984strange}, and since then the study of it has become an active area of research in nonlinear dynamics. These attractors are realized in many theoretical models and experiments. Apart from observations of these attractors in typical nonlinear systems such as logistic map, circle map, Duffing oscillator and van der Pol oscillator \cite{tomasz1994attractors,ulrike2006strange} under quasiperiodic forcing, these exotic attractors have also been experimentally observed in a quasiperiodically driven magnetoelastic ribbon system \cite{ditto1990experimental}, in electronic circuits \cite{zhou1992observation, thamilmaran2006experimental}, in a plasma system \cite{ding1997observation}, in an electrochemical cell \cite{ruiz2007experimental} and in a system near the torus-doubling critical point \cite{bezruchko2000experimental}. The evidence for these attractors have been recently realized in the pulsation of stars like KIC 5520878 \cite{lindner2015strange, lindner2016simple}, in a nonsmooth dynamical system\cite{li2019strange}, a cellular neural network\cite{megavarna2019observation}, and the quasiperiodically driven geophysical Saltzman model\cite{middya2019parametrically}. 

The study of  SNA is of particular physical interest in a quantum particle in a spatially quasiperiodic potential\cite{bondeson1985quasiperiodically}. SNAs are important in biological systems too \cite{ding1994phase, prasad2003strange} and for communication as well\cite{zhou1997robust,ramaswamy1997synchronization}.   	
Many of the previous studies focused on the routes and mechanisms by which an SNA is generated from a regular attractor or disappears as a chaotic attractor  \cite{heagy1994birth, nishikawa1996fractalization, yalccinkaya1996blowout,  prasad1997prasad,  venkatesan1999birth, lai1996transition, witt1997birth, prasad1999collision, wang2004strange, venkatesan2000intermittency, prasad2001strange, venkatesan2001interruption, gopal2013applicability}. Different routes to SNAs have been established in different nonlinear dynamical systems. In particular, routes like Heagy-Hammel route \cite{heagy1994birth}, fractilization of torus\cite{nishikawa1996fractalization}, blowout bifurcation\cite{yalccinkaya1996blowout} and intermittency \cite{prasad1997prasad,venkatesan1999birth} and other routes have been reported \cite{lai1996transition,witt1997birth,prasad1999collision,wang2004strange,venkatesan2000intermittency,prasad2001strange,venkatesan2001interruption,gopal2013applicability}.   
SNAs are also characterized by various tools including finite time Lyapunov exponents, phase sensitivity exponents, 0-1 test, recurrence plots, spectral distribution and so on  \cite{lindner2015strange, lindner2016simple, heagy1994birth, nishikawa1996fractalization, yalccinkaya1996blowout, lai1996transition, prasad1997prasad, witt1997birth, prasad1999collision, wang2004strange, venkatesan2000intermittency, prasad2001strange, venkatesan1999birth, venkatesan2001interruption, gopal2013applicability, pikovsky1995singular, pikovsky1995characterizing}. Mathematically related issues have also been addressed corresponding to the generation and properties of SNA \cite{stark1997invariant, sturman2000semi}.

Since many of the physically realized nonlinear dynamical systems do not fall under the category of quasiperiodic forcing, it is natural to question whether it is possible to realize SNAs in nonlinear systems with other types of forcings \cite{wang2006characterization}. It was shown in the literature that two asymmetrically coupled driven ring maps and a periodically driven oscillator with an inertial nonlinearity can produce SNAs via band merging crisis \cite{anishchenko1996strange}. Later, it was proved that these attractors are actually chaotic \cite{pikovsky1997comment}.  Aperiodic nonchaotic attractors have been created in nonlinear systems using periodic forcing of high period \cite{nandi2009design}. Many works have also focused on the generation of SNAs via stochastic forcing \cite{rajasekar1995controlling,prasad1999can,wang2004strange,sturman2000semi,wang2006characterization}. It was suggested that a chaotic attractor can be converted into an SNA by adding suitable noise\cite{rajasekar1995controlling}. Later it was proved that the effect of noise smears out the fine structure of the attractor and gives rise to negative Lyapunov exponents. Thus, the dynamics of this case is neither strange nor nonchaotic\cite{prasad1999can}. Wang \textit{et al.} reported that robust SNAs can be induced by noise in autonomous discrete-maps and in periodically driven continuous systems \cite{wang2004strange, wang2006characterization}. They have shown that in the periodic window, if the strength of noise satisfies the condition that $ D>D_{m} $, a critical value, the trajectory of the system switches intermittently between the periodic attractor and the chaotic saddle \cite{wang2006characterization}. The Lyapunov exponent remains negative for noise amplitude $ D_{m}<D<D_{*m} $, where $ D_{*m} $ is the noise amplitude for which the maximal Lyapunov exponent is zero. In this range, the attractor of the system has a strange geometry but the maximal Lyapunov exponent is non-positive \cite{wang2004strange, wang2006characterization}. They termed this attractor as a noise-induced strange nonchaotic attractor. Further, such SNAs have occurred in a small region of the parameter space.

Thus, a very basic question arises : can it be shown that SNAs exist in dynamical systems other than quasiperiodically forced ones and whether they are robust? Here robust SNA is referred to the attractor to which an arbitrarily small change of the system can not cause its destruction\cite{hunt2001fractal,kim2003fractal}. Robustness of SNA is essentially connected to its observation in experiments. Hunt and Ott have established analytically that in a simple mapping robust SNAs are generic under quasiperiodic forcing \cite{sturman2000semi, hunt2001fractal}. They have shown that such robust strangeness of the attractors can be verified by calculating the information dimension (which has the value `1') and capacity dimension (which should be two for the mapping considered). Such an analysis is employed for a few more examples \cite{hunt2001fractal,kim2003fractal}. Most of the studies of SNAs relied on computational verification. Even numerical approach has its own limitations. It was pointed out that the finite precision of calculation of computer leads to numerical errors \cite{shi2008relation}. 
 
In the present paper, we address these issues by examining the behavior of a periodically driven nonlinear oscillator in the presence of noise. Specifically, we report the existence of SNA in an optimal range of noise strength in a periodically driven double-well Duffing oscillator with a single force. To validate the robustness of SNA, we perturb the system with logical signals and examine whether this perturbation can alter the existence of SNAs in the system. We find that the SNAs persist  in the noise induced periodically driven nonlinear system, even under perturbation.  

It is well known that noise is inevitable in many physical situations and in fact it puts an upper limit on the performance of the system. In particular, the factor of noise is the main concern as well as the limiting factor in the designing of digital integrated circuits and ultimately computer architecture. Although many nonlinear dynamical computing systems have been proposed to make computing robust, reliable and reconfigurable \cite{sinha1998dynamics, sinha1999computing,murali2007using,murali2009reliable,kia2011unstable, dari2011creating,kohar2017implementing,kia2017nonlinear}, the ambient noise and practical nonidealities restrict one to emulate different logic elements. In this regard, recently the present authors have demonstrated a route to logical SNA in quasiperiodically driven nonlinear systems \cite{aravindh2018strange}. They showed that if the quasiperiodically driven Duffing oscillator were perturbed by two logic signals, the output of the system reproduces logical behavior. They further demonstrated that by using such robust SNAs, one can emulate different logic gates in the presence of noise \cite{aravindh2018strange}.

In principle, it is possible to generate and maintain quasiperiodicity in a simple way  but in practice it is difficult to carry this out. For quasiperiodicity, forcing may be given with an irrational frequency or can be generated with two sources whose frequencies are incommensurate. Experimental uncertainties can usually lead to deviations in the precision of measuring rational or irrational numbers. Thus a question arises naturally: Can logical SNA arise in the absence of quasiperiodic forcing? In other words whether logical SNAs can arise in situations where the underlying system has other forcing dependencies. In the present work, we identify a route to logical SNA induced by noise in periodically driven Duffing oscillator with a single force. We perturb the system by using two square waves in the presence of noise and establish that the perturbed attractor persists with SNA properties. We further show that the output of the system reproduces logic elements controlled by noise. When we change the threshold or biasing of the system, the response of the system changes from one logical operation to another one. 
		
The paper is organized as follows.  We discuss the basic aspects of periodically driven Duffing oscillator in Sec.II. We present noise induced SNAs in the above periodically driven system in Sec.III. We also describe the route to logical SNA and deduce the effect of two logical inputs. We further discuss how to deduce the probability of getting logical behavior and then show how to implement different logical behavior in Sec.IV. Finally, we summarize our results in Sec.V.
	
\section{Periodically driven double-well Duffing oscillator}
In the present work, a periodically driven double-well Duffing oscillator by a single sinusoidal force of the following form is considered:
	\begin{align}
		\dot{x}&=y,  \nonumber \\
		\dot{y}&=-\alpha \dot{x} - \beta (x^3-x)+ F\sin \theta + \varepsilon+\sqrt{D}\xi(t)+I, \nonumber \\ 
		\dot{\theta}&= \omega.
		\label{equ1}
	\end{align}         

	Eq.\eqref{equ1} is assumed as the equation of motion for a particle of unit mass in the potential well\cite{lakshmanan1996chaos} $ V(x)=\beta\big(-\dfrac{1}{2}x^{2}+\dfrac{1}{4}x^{4} \big) $. The simplest experimental realization of Eq.\eqref{equ1} is a magnetoelastic ribbon \cite{ditto1990experimental}.

	The quantities $F$ and $\omega$  in (\ref{equ1}) are the  amplitude and frequency of the external forcing, respectively. $\varepsilon$ and $\xi(t)$ correspond to the asymmetric bias constant input and Gaussian white noise of intensity $D$, respectively. $I$ is the amplitude of the input square wave signal. In our study we vary `$F$', the amplitude of forcing parameter. The parameters are fixed for our numerical calculations as $\alpha=0.5,~ \beta=1.0,~\omega=1.0$  and $ \varepsilon=0.1 $ in  Eq.\eqref{equ1}. It is an established fact that the system \eqref{equ1} exhibits period-doubling route to chaos, intermittency route and so on \cite{venkatesan2000intermittency,lakshmanan2003chaos} in the absence of noise $ (D=0.0) $.
	
	\section{Noise-induced SNAs in periodically driven Duffing oscillator}
	
\begin{figure}[h!]
		\centering
		\includegraphics[width=0.8\linewidth]{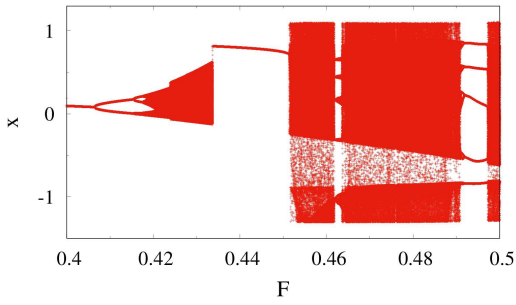}
		\caption{Bifurcation diagram of \emph{F} vs x(t) in the presence of bias $ \varepsilon=0.1 $ and absence of logic inputs intensity $ \delta = 0.0 $ and Gaussian white noise $ D=0.0 $.}
		\label{fig1}
	\end{figure}

	In the absence of noise $ (D=0) $, bias $ \varepsilon=0.1 $ and logic input $ (I=0) $, we vary the forcing parameter `$ F $'. It is observed that the system \eqref{equ1} exhibits typical period-doubling route to chaos as shown in Fig.\ref{fig1}.
	
\begin{figure}
	\centering
	\includegraphics[width=0.323\linewidth]{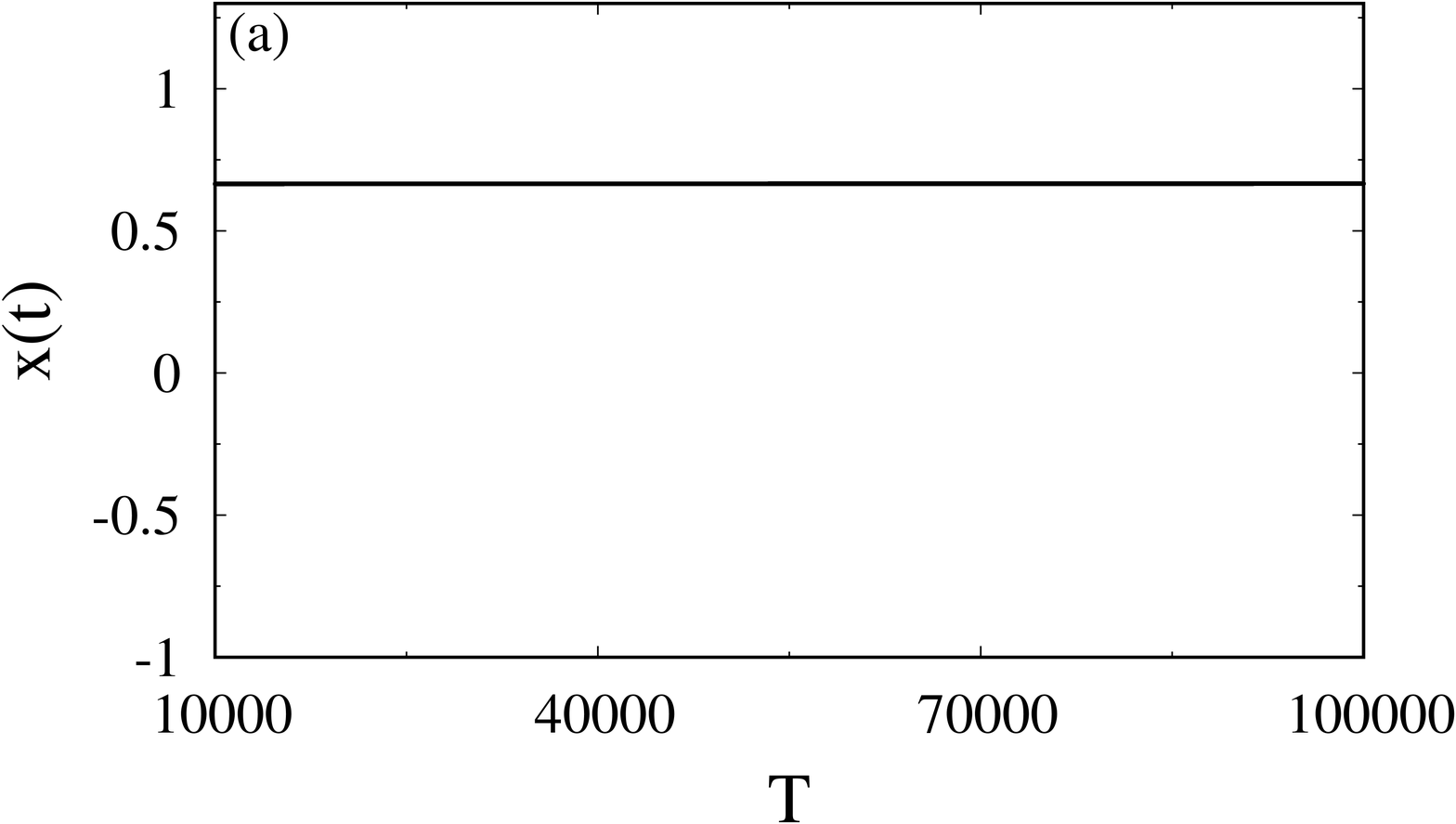}
	\includegraphics[width=0.323\linewidth]{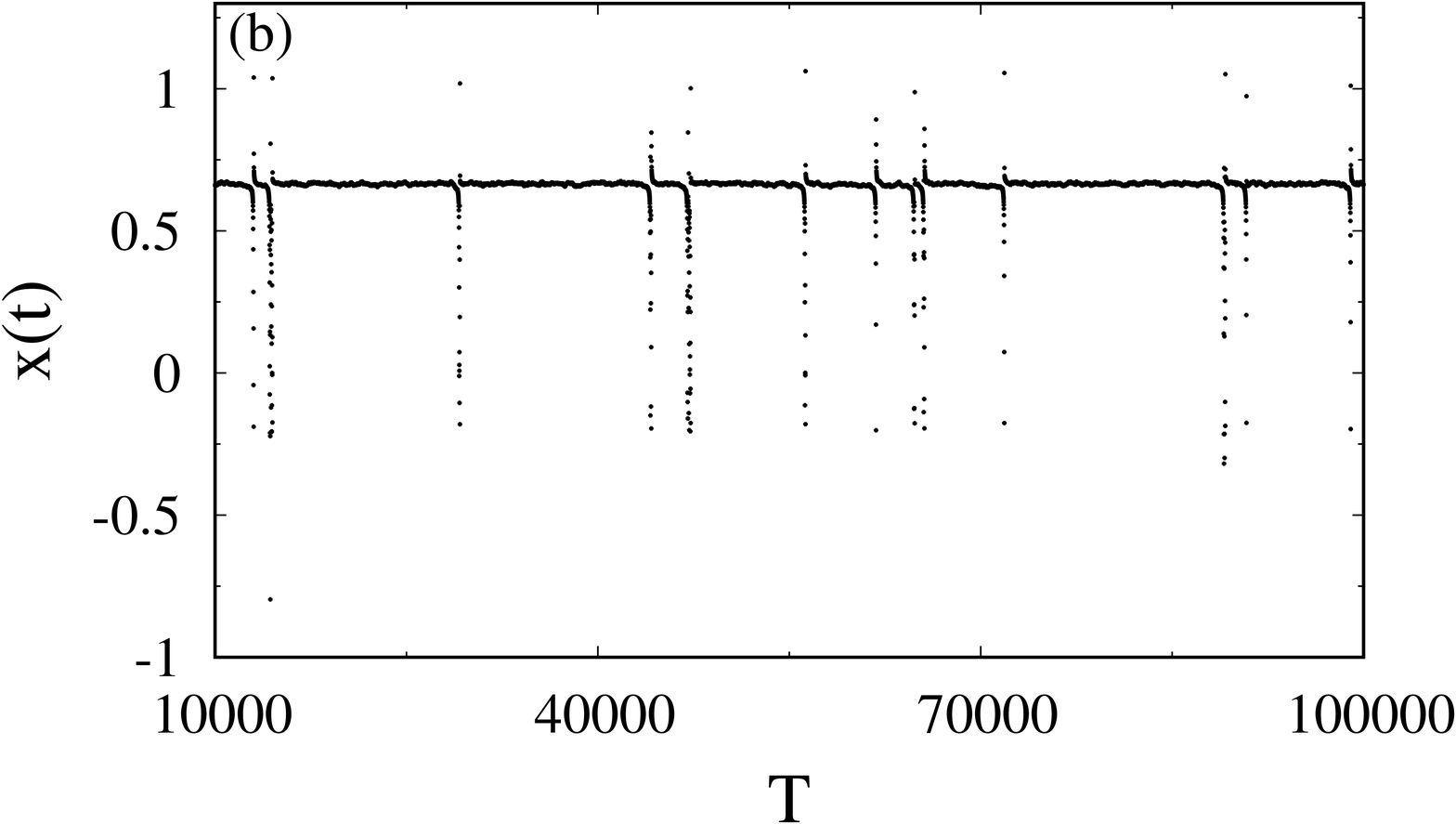}
	\includegraphics[width=0.323\linewidth]{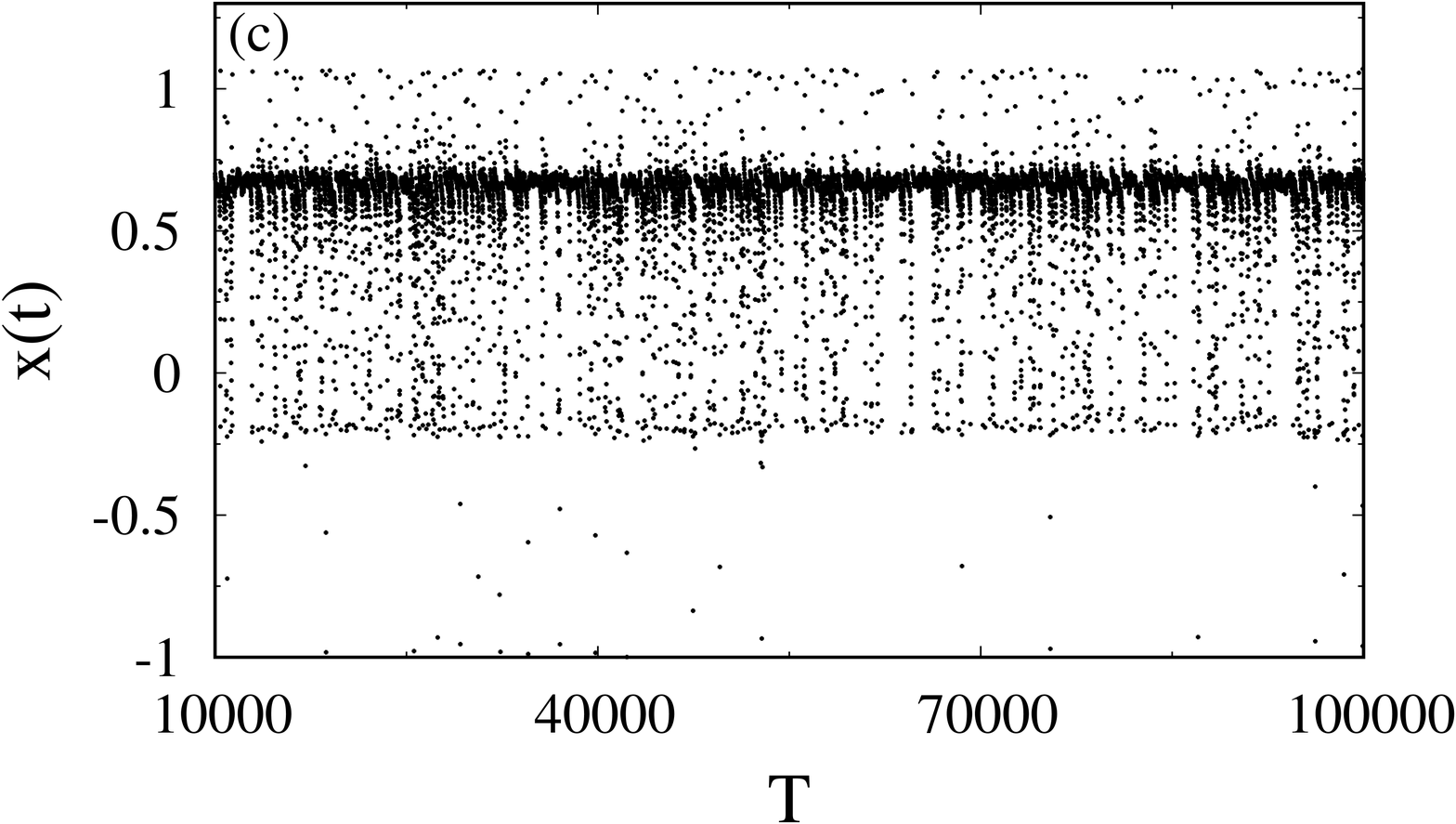}
	\caption{Panels (a)-(c) show the Poincar\`e surface of section in the time series plane with different values of noise strength with  (a) $ D=0.0 $ , (b) $ D=0.00001 $ and (c) $ D=0.0001 $ with fixed forcing parameter value $ F=0.4514 $ and bias $ \varepsilon=0.1 $.}
	\label{fig15}
	\end{figure}

\begin{figure}
		\centering
		\includegraphics[width=0.8\linewidth]{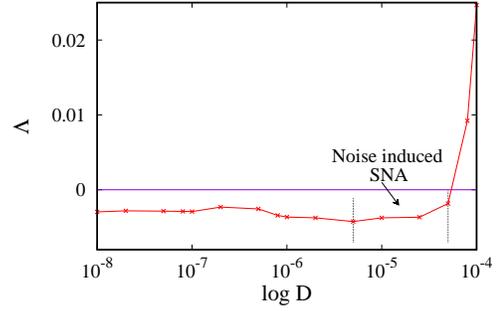}
		\caption{Largest Lyapunov exponent $ \Lambda $ against the noise amplitude `D'.}
		\label{fig7}
	\end{figure}

	Now, we consider the periodic window in Fig.\ref{fig1} in the range $ 0.43365<F<0.45164 $. Let $ F_{m} $ and $ F_{m}^{*} $ be the parameter values at the beginning and end of the periodic window where the maximal Lyapunov exponent is negative.  It is well known that at the end of the window a chaotic saddle coexists with periodic orbits and this leads to transient chaos \cite{wang2004strange,wang2006characterization}. Now we fix the system parameter value at $ F=0.4514 $, where the system behaves periodically[see Fig.\ref{fig15}(a)] and the corresponding largest Lyapunov exponent (LE) is $ \lambda=-0.0028 $ [see Fig.\ref{fig7}]. Now, by applying external noise with amplitude $ D>D_{m} = 0.00001$, the periodic attractor can be made to disappear. Here, the trajectory of the system switches intermittently between the periodic attractor and chaotic saddle as shown in the time trajectory plot in Fig.\ref{fig15}(b).  As a result, the structure of the asymptotic attractor changes gradually [see Fig.\ref{fig15}(b)]. Though the attractor encompasses the components of periodic and chaotic behaviors, the largest Lyapunov exponent is still negative [see Fig.\ref{fig7}] for the noise strengths lying between the values $ D_{m}=0.000005 < D<D_{m}^{*}=0.00005$ and this attractor is considered as an SNA \cite{wang2004strange} [see Figs.\ref{fig15}(b) for $ D=0.00001 $]. On further increase of $ D>0.00005 $ the system behaves as a chaotic one [see Fig.\ref{fig15}(c)] and the corresponding largest Lyapunov exponent becomes positive as shown in Fig.\ref{fig7}.
	
\begin{figure}
		\centering	
		\includegraphics[width=0.32\linewidth]{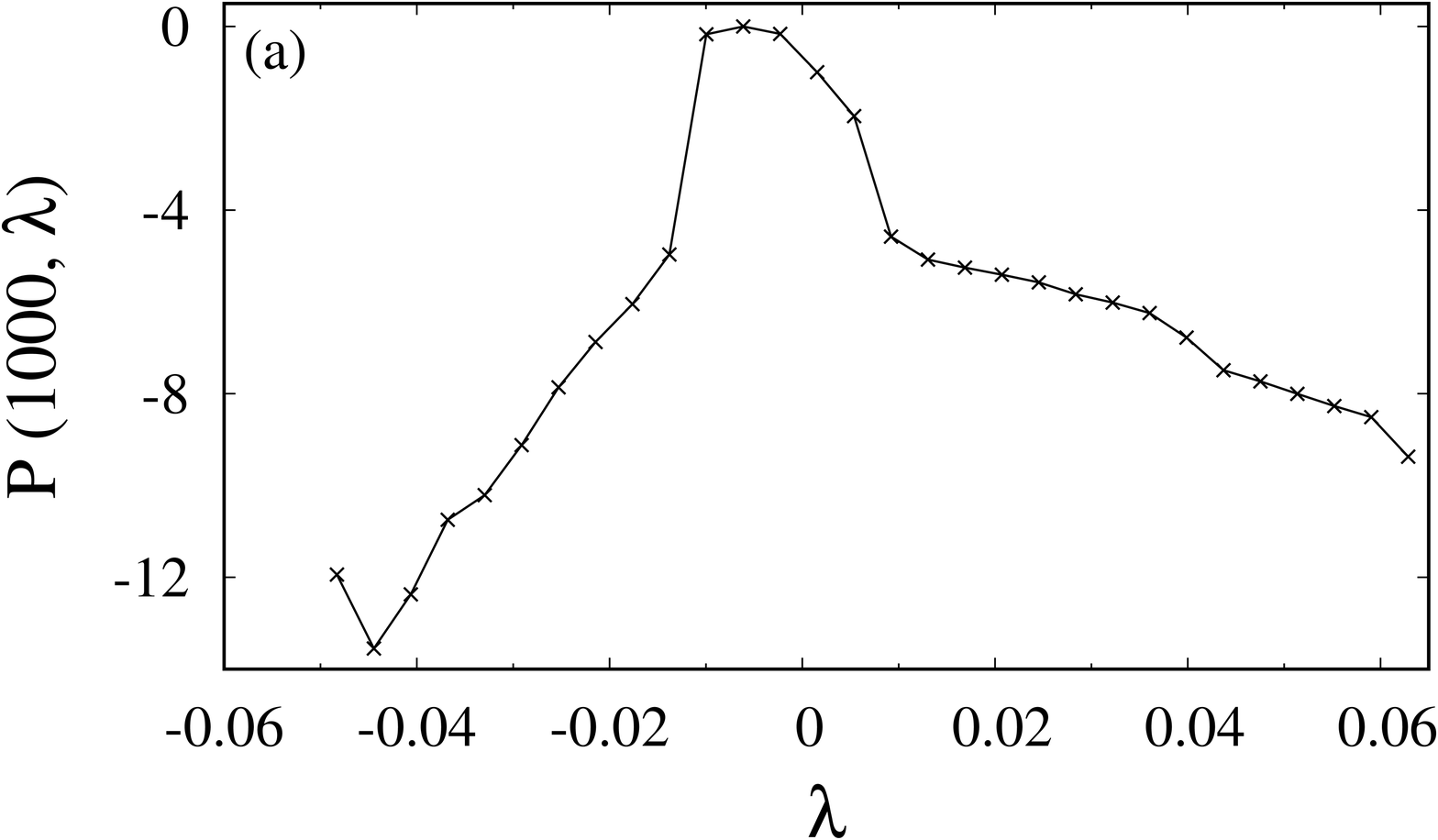}
		\includegraphics[width=0.32\linewidth]{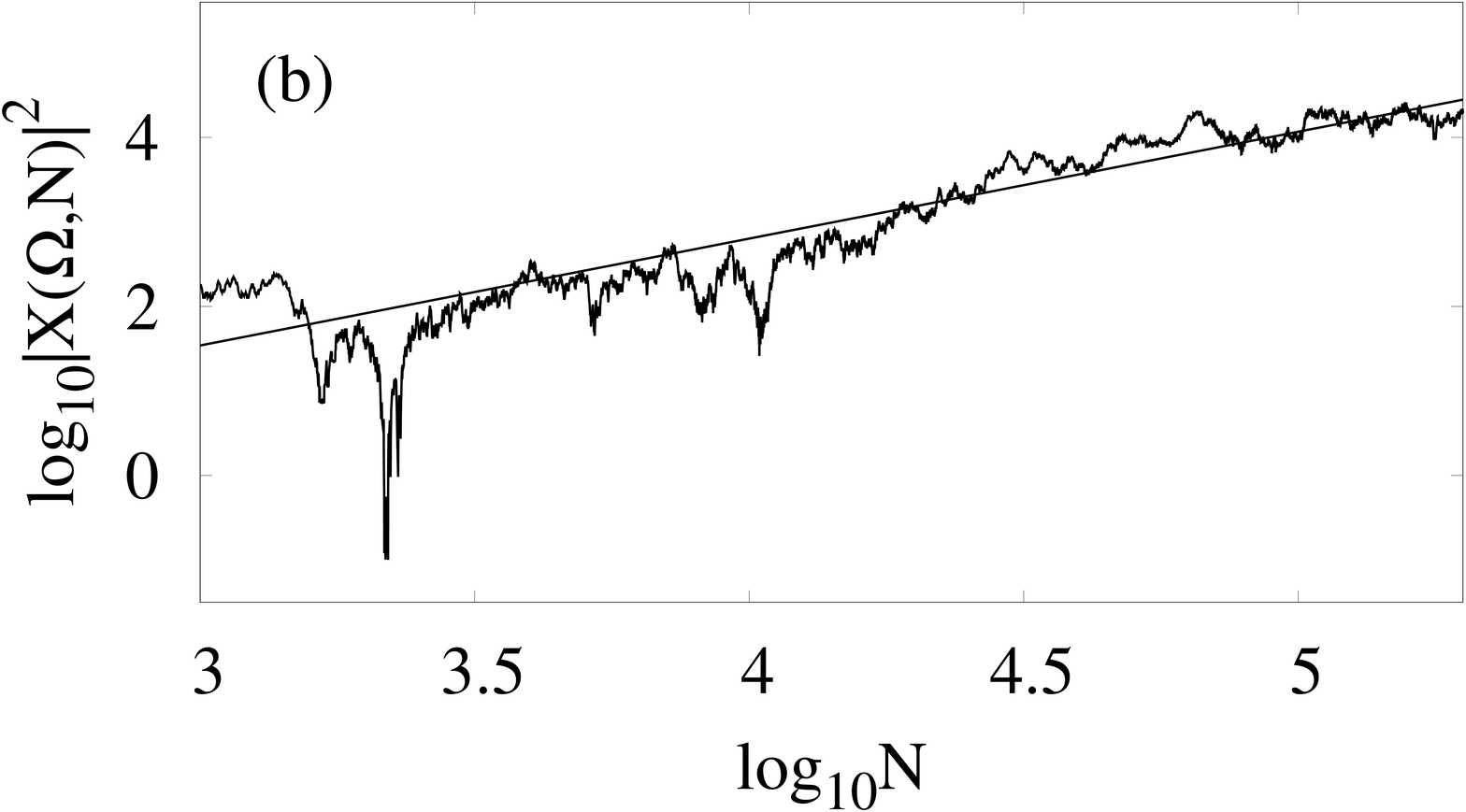}
		\includegraphics[width=0.32\linewidth]{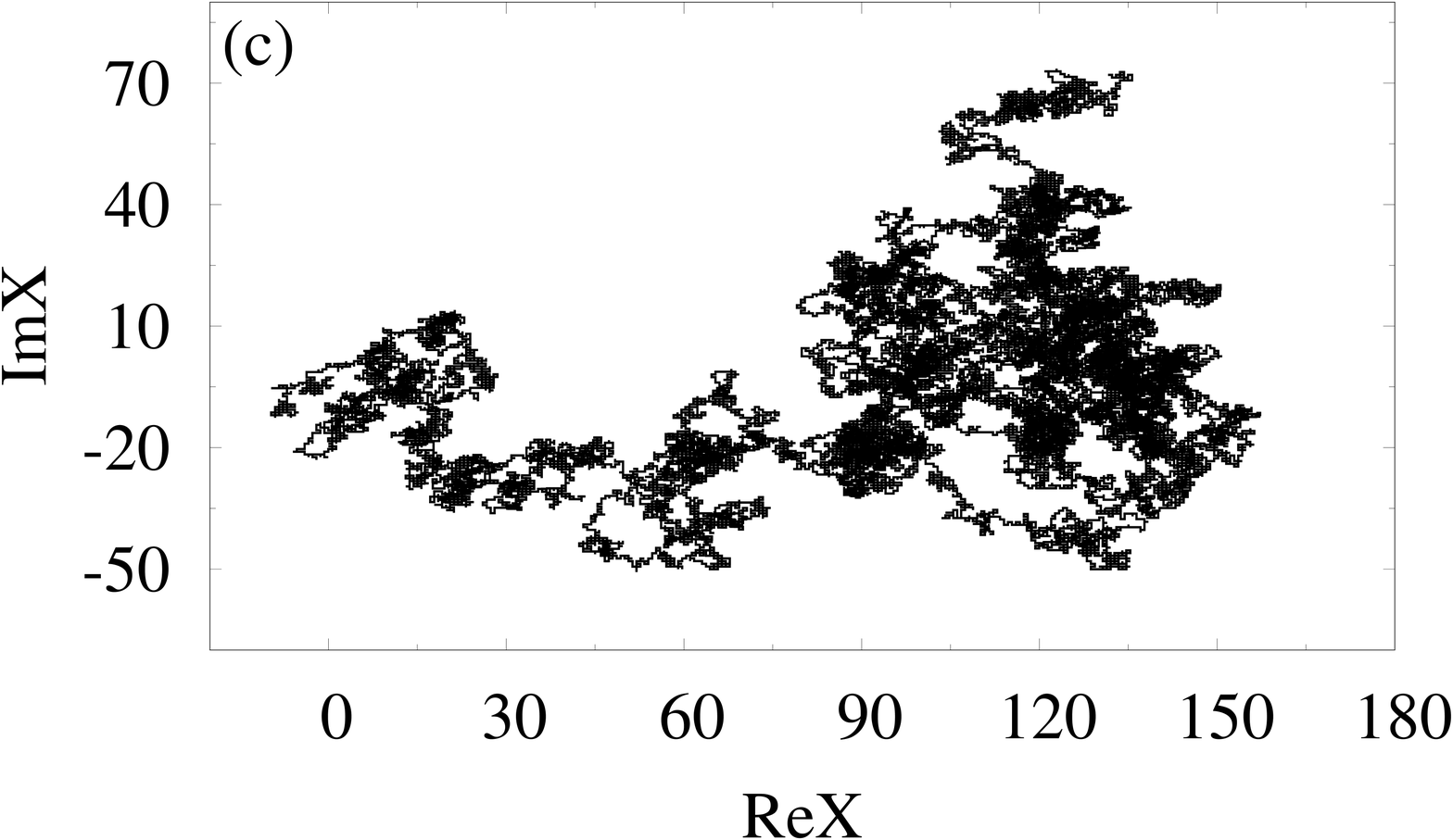} 
		\caption{Panel (a) shows the distribution of the finite-time Lyapunov exponent in the presence of noise with $ D=0.00001 $. Panel (b) shows the Fourier spectrum $|X(\Omega,N)|^{2}$ vs $N^{\gamma}$ on logarithmic scale with $\gamma=1.26454 $ for SNA. Panel (c) represents the fractal nature of trajectories in the complex plane. The fixed parameters are $ F=0.4514 $, bias $ \varepsilon=0.1 $ and in the absence of logic input intensity $ \delta=0.0 $.}
		\label{fig2a}
	\end{figure}

	We confirm that a suitable noise can induce the SNA in the periodically driven system with the help of Poincar\`e surface of section, distribution of finite time Lyapunov exponents (FTLE) and Fourier power spectra, which are shown in Figs.\ref{fig2a}. In Figs.\ref{fig2a}(a) we depict the distribution of the finite time Lyapunov exponent (FTLE) in the presence of noise. Here it is found that the exponent is mostly in the negative side but it intermittently appears in the positive side too and is apparently exponential [see Fig.\ref{fig2a}(a)]. In Figs.\ref{fig2a}(b) \& \ref{fig2a}(c) we show the spectral properties and fractal walk. 
 	It is evident from spectral distribution that the power spectrum obeys the scaling relation $|X(\Omega,N)|^{2} \sim N^{\gamma}$,  where $\gamma=1.26454$ for the SNA [see Fig.\ref{fig2a}(b)]. Further, the trajectories in the complex plane of \emph{(ReX, ImX)} exhibit fractal behavior as shown in Fig.\ref{fig2a}(c), for fixed parameters $ F=0.4514 $, $ D=0.00001 $ and with bias $\varepsilon=0.1$,  and  in the absence of logic input $ \delta=0.0 $ .

	\begin{figure}[h!]
		\centering
		\includegraphics[width=0.48\linewidth]{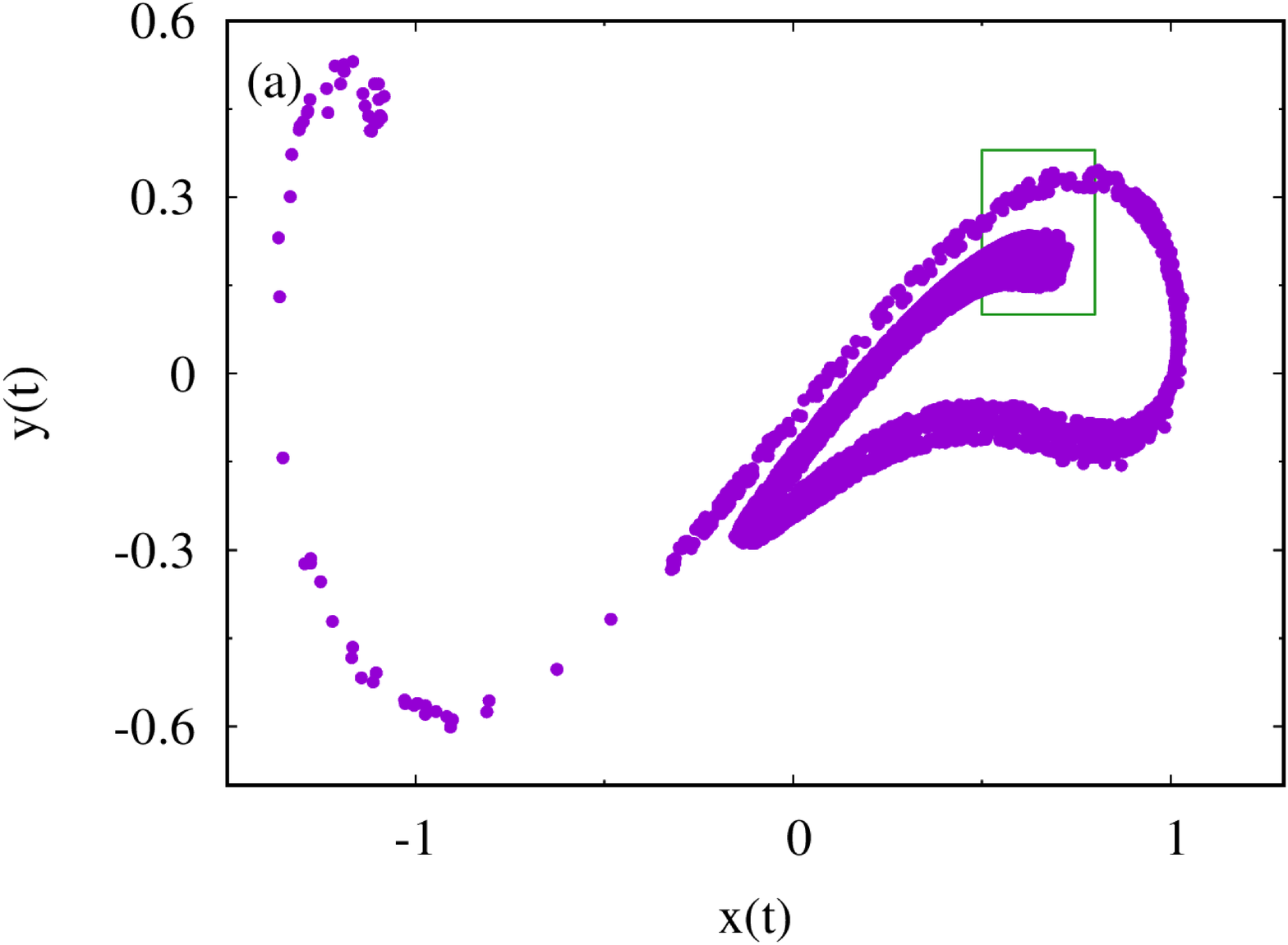}
		\includegraphics[width=0.48\linewidth]{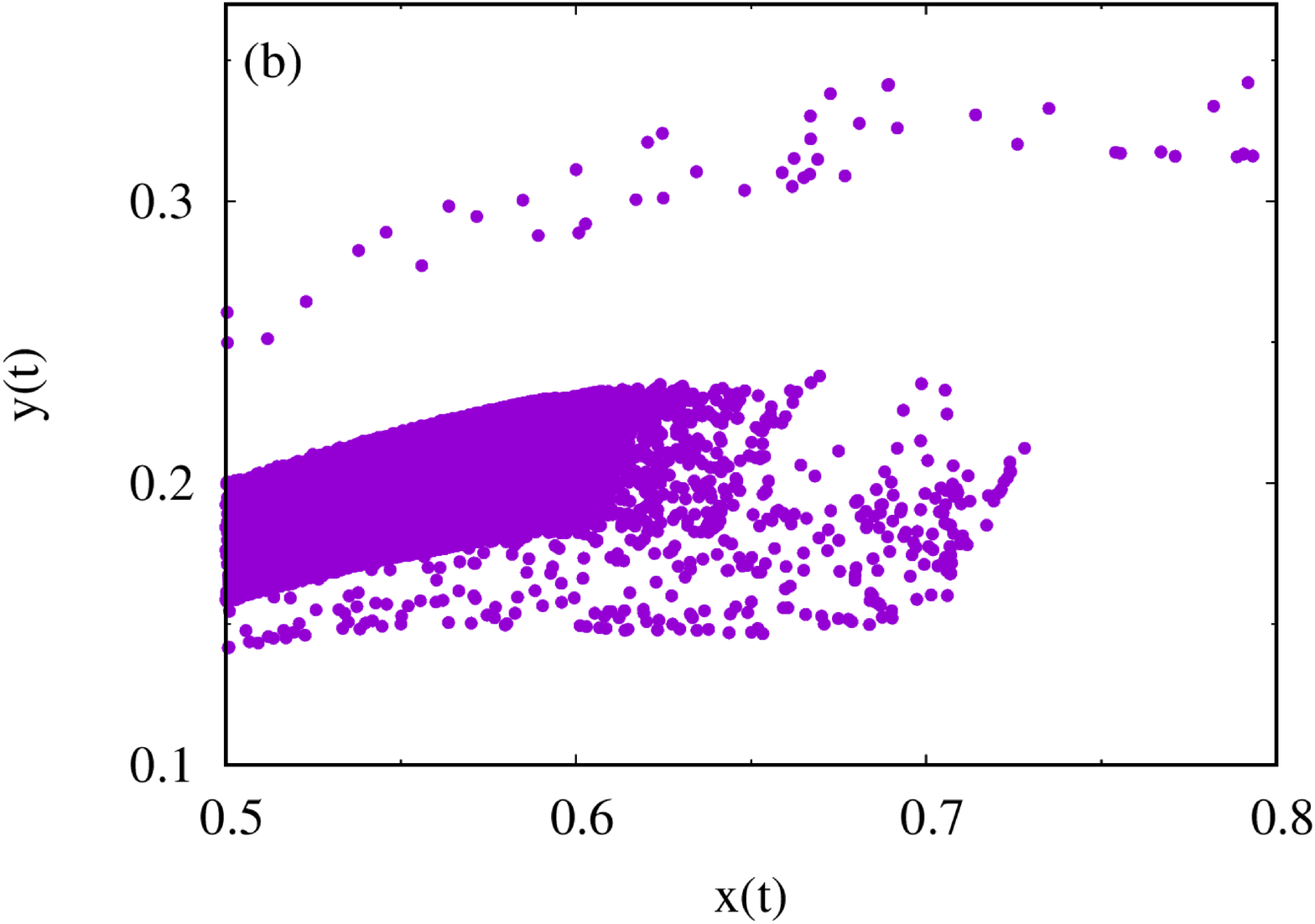}
		\includegraphics[width=0.48\linewidth]{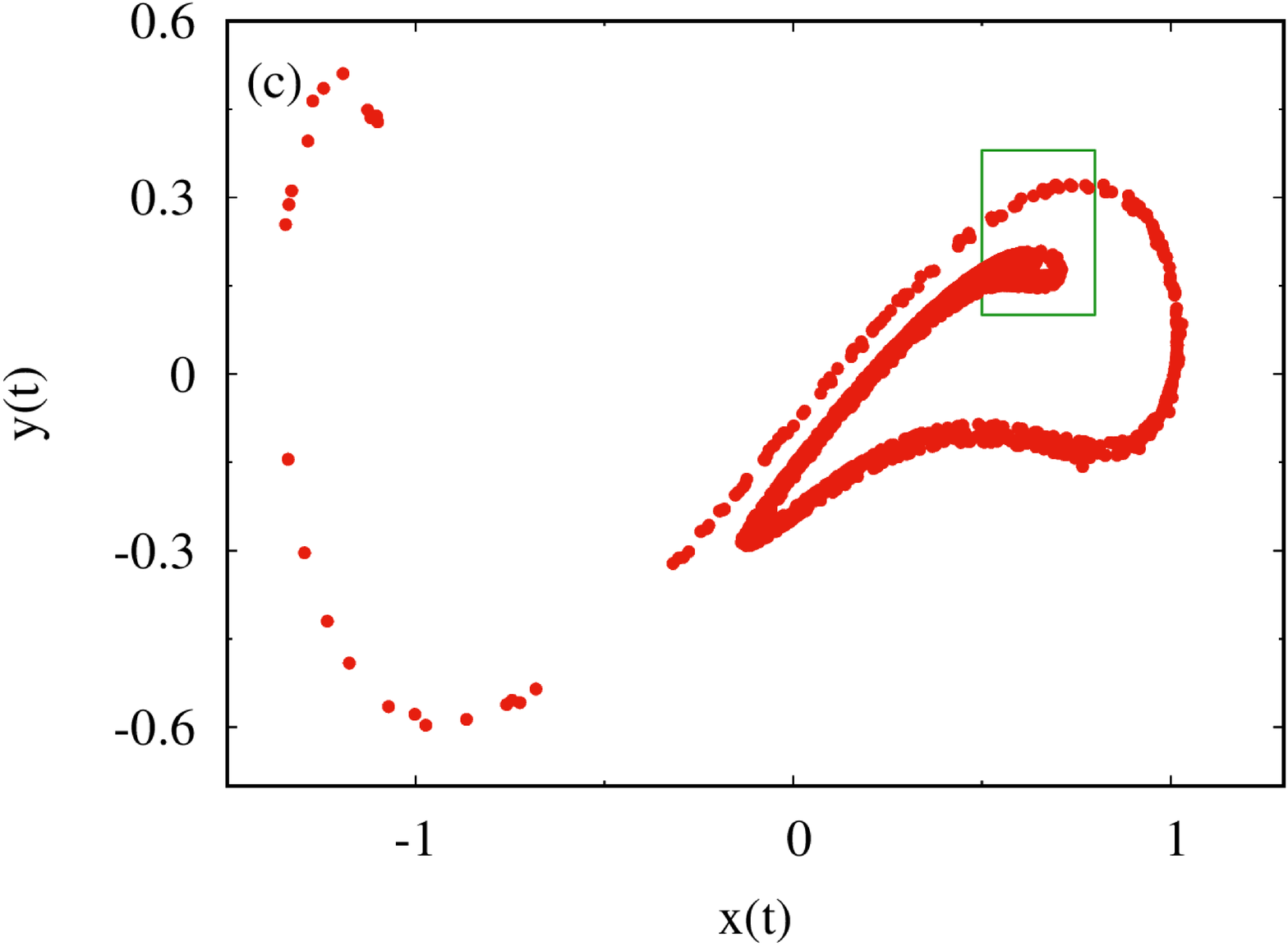}
		\includegraphics[width=0.48\linewidth]{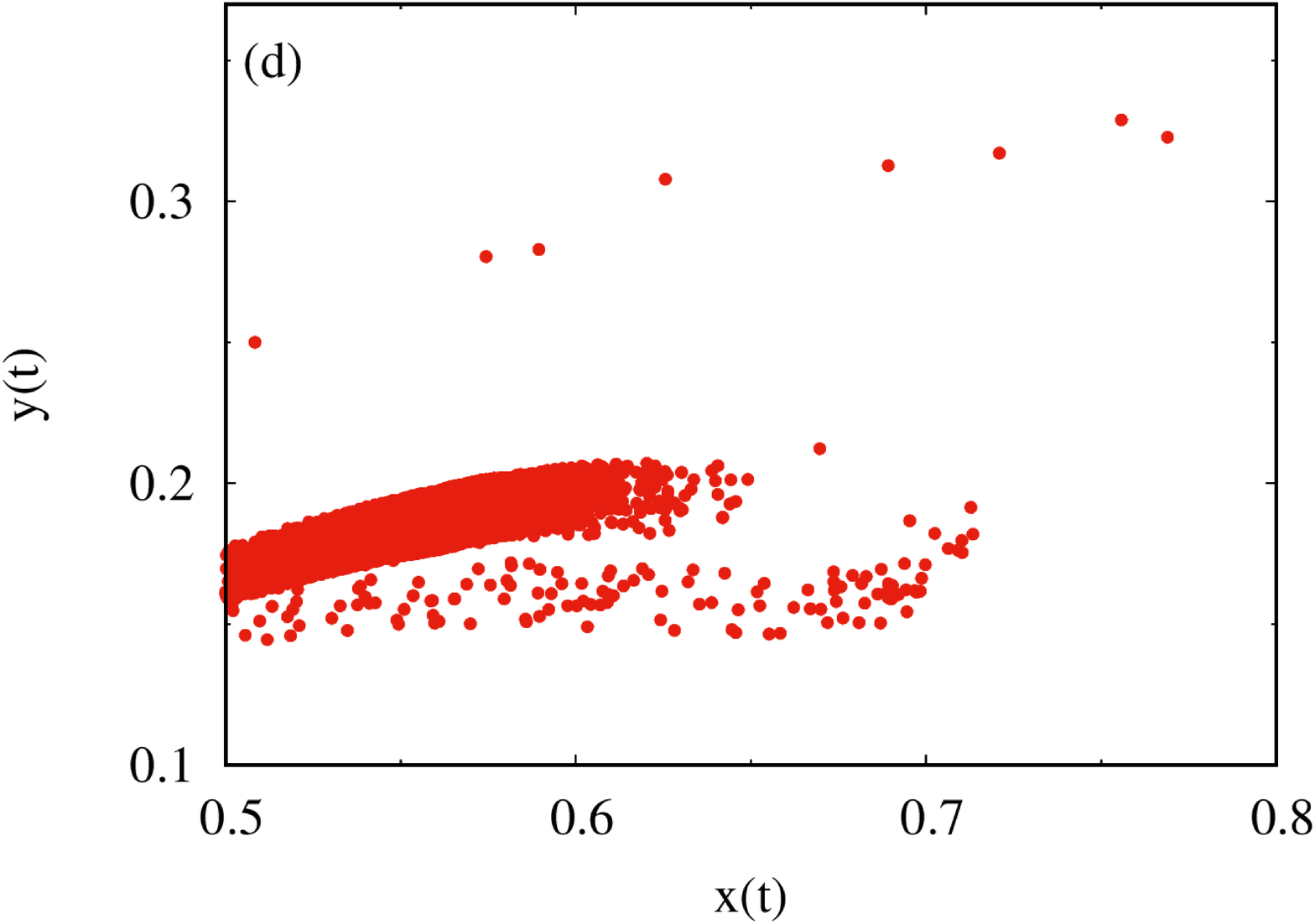}
		\caption{Poincar\'e section of the phase space under white noise of amplitude $ D=0.00005 $ with bias $ \varepsilon=0.1 $ and without logic pulse:-  Panels (a) \& (b) respectively represent a single trajectory and magnified part of it, and Panels (c) \& (d) respectively correspond to the snapshot attractors formed from out of 10,000 trajectories and a blow-up part of (c).}
		\label{fig8}
	\end{figure}

	It is well known that noise can smear out any strange geometry of the underlying attractor in random dynamical systems. Romeiras \textit{et al.} have pointed out that the fractal structure of chaotic attractor  can be resolved even under noise by analyzing the snapshot attractors constructed from out of a large number of trajectories \cite{romeiras1990multifractal}. Wang \textit{et al.} have explored these snapshot attractors for resolving the strange geometry of noise induced SNAs \cite{wang2004strange, wang2006characterization}. Following their work, we also examine the snapshot attractors for the present system formed by a large number of trajectories. In particular we obtain the snapshot attractors by using a grid of $ 100 \times 100 $ initial conditions uniformly distributed in the region $ -2.0 \le (x_{0}, y_{0}) \le 2.0 $ and the amplitude of the inhomogeneous noise is $ 10^{-10} $. Figs.\ref{fig8}(a) and \ref{fig8}(b) represent, respectively, a single trajectory and its blow-up. Here, the points of the single trajectory and its blow-up part show that the points are randomly distributed. However, the snapshot attractors formed by 10,000 trajectories as shown in fig. \ref{fig8}(c) and it blow-up part as shown in Fig.\ref{fig8}(d) exhibit  apparently a  fractal structure.

 \section{Route to logical SNA with single periodic force and noise}
	
\subsection{Effect of three level square wave in noise induced SNA}

Past studies revealed that SNAs are typical in quasiperiodically driven nonlinear dynamical systems, and it is important to inquire whether the behavior of SNA in the present case is robust. When we say that the SNA is robust it means that the behavior persists under sufficiently small perturbations. It is an established fact that all robust behaviors are also typical, however not vice-versa \cite{hunt2001fractal,kim2003fractal}. Motivated by the above facts, we investigate whether the SNA observed in noisy, periodically driven nonlinear system is typical and robust. For this purpose, we perturb the system with logic signals and analyze whether SNA persists or not. 
	
In particular, we analyze the response of the system \eqref{equ1} under the effect of a logic input signal $ I $. Specifically, the system \eqref{equ1} is driven with a low/moderate amplitude logic input signal $ I $, where, $I=I_{1}+I_{2}$ with two square waves of strengths $I_1$ and $I_2$ encoding two logic inputs. The inputs can be either 1 or 0, giving rise to four distinct logic input  sets $(I_1,I_2):(0,0),(0,1),(1,0)$ and $(1,1)$. For a logical  `$1$', we set $I_{1}=I_{2}=+\delta$, while for a `$0$', we  set $I_{1}=I_{2}=-\delta$. Here $ \delta $ represents the strength of the input signal. We also note that the input sets (0,1) and (1,0) correspond to the same input signal $I$. As a result, the four distinct input combinations $(I_1,I_2)$ reduce to three distinct values of $I$, namely $-2\delta,~ 0,~ +2\delta$, corresponding to the logic inputs $(0,0),~(0,1)$ or $(1,0),~(1,1)$,  respectively.

	\begin{figure}[h!]
		\centering	
		\includegraphics[width=0.8\linewidth]{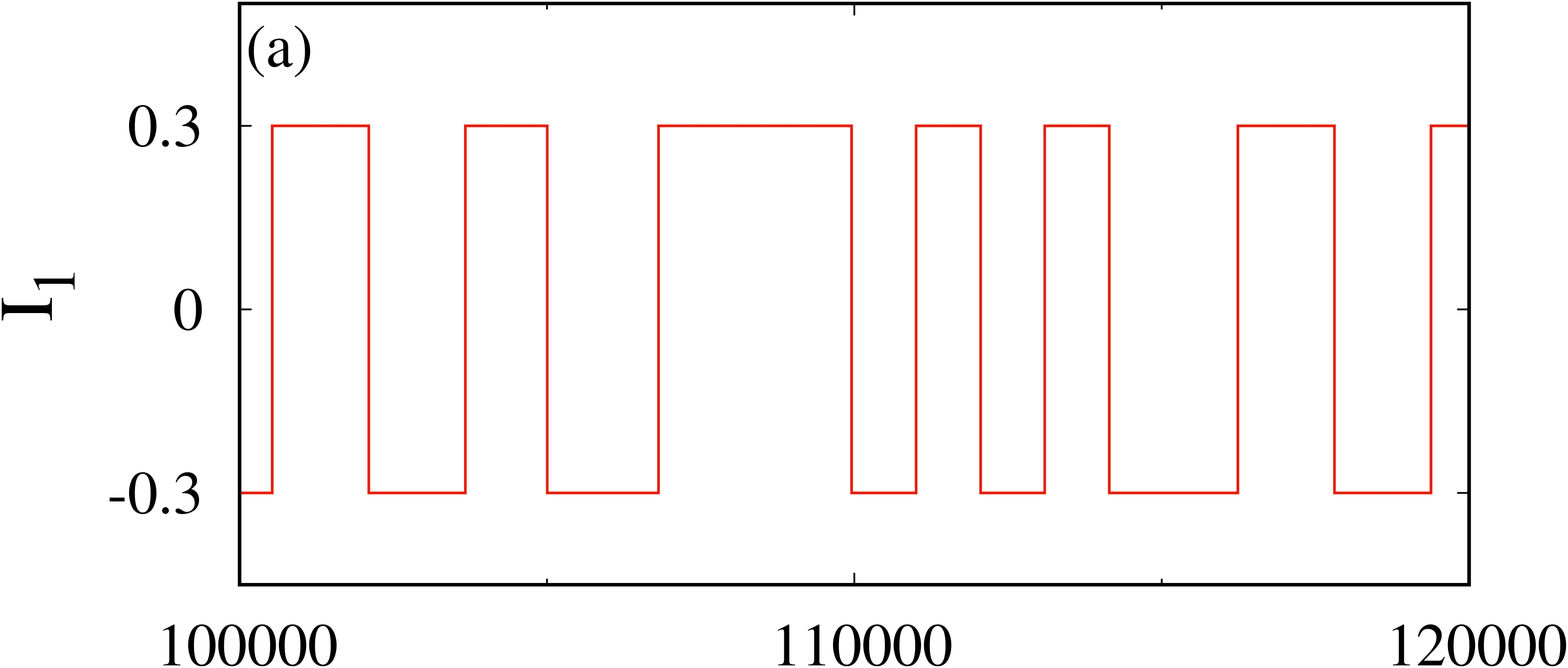} 		
		\includegraphics[width=0.8\linewidth]{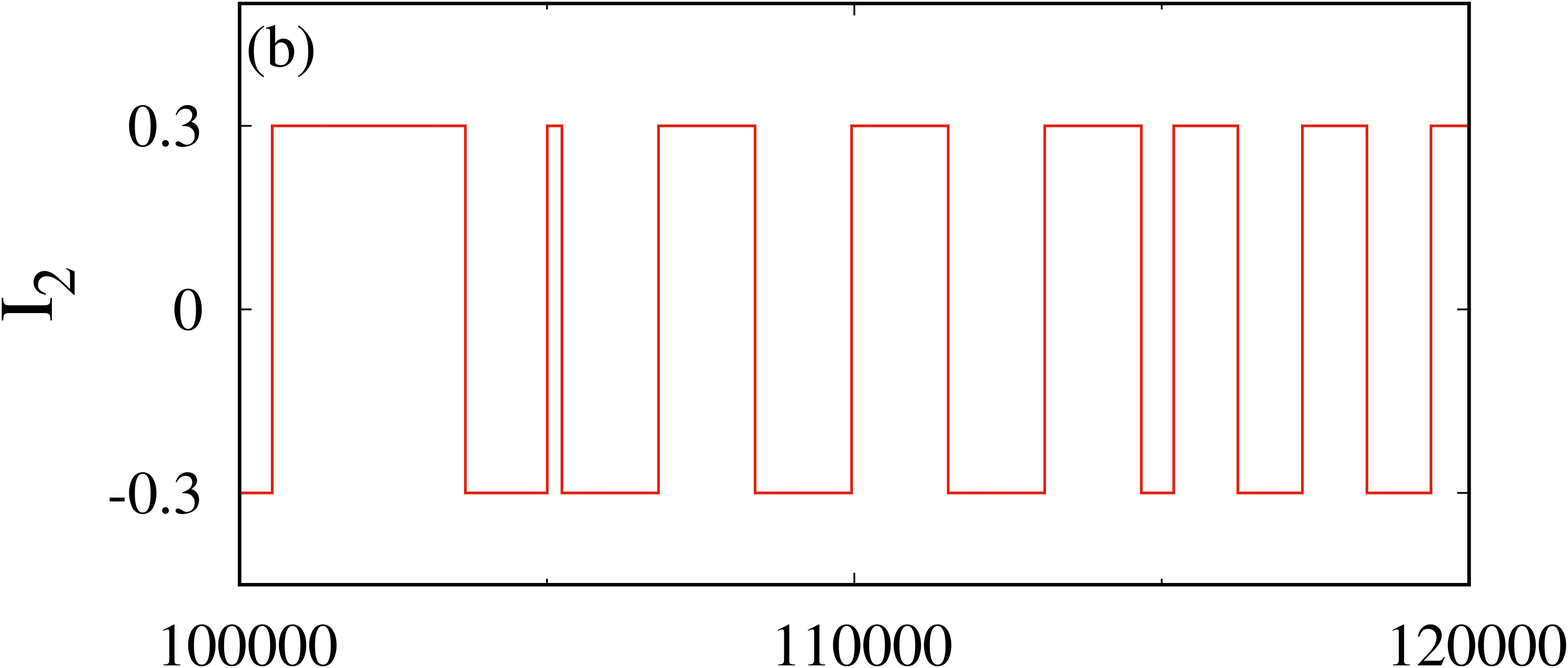} 
		\includegraphics[width=0.8\linewidth]{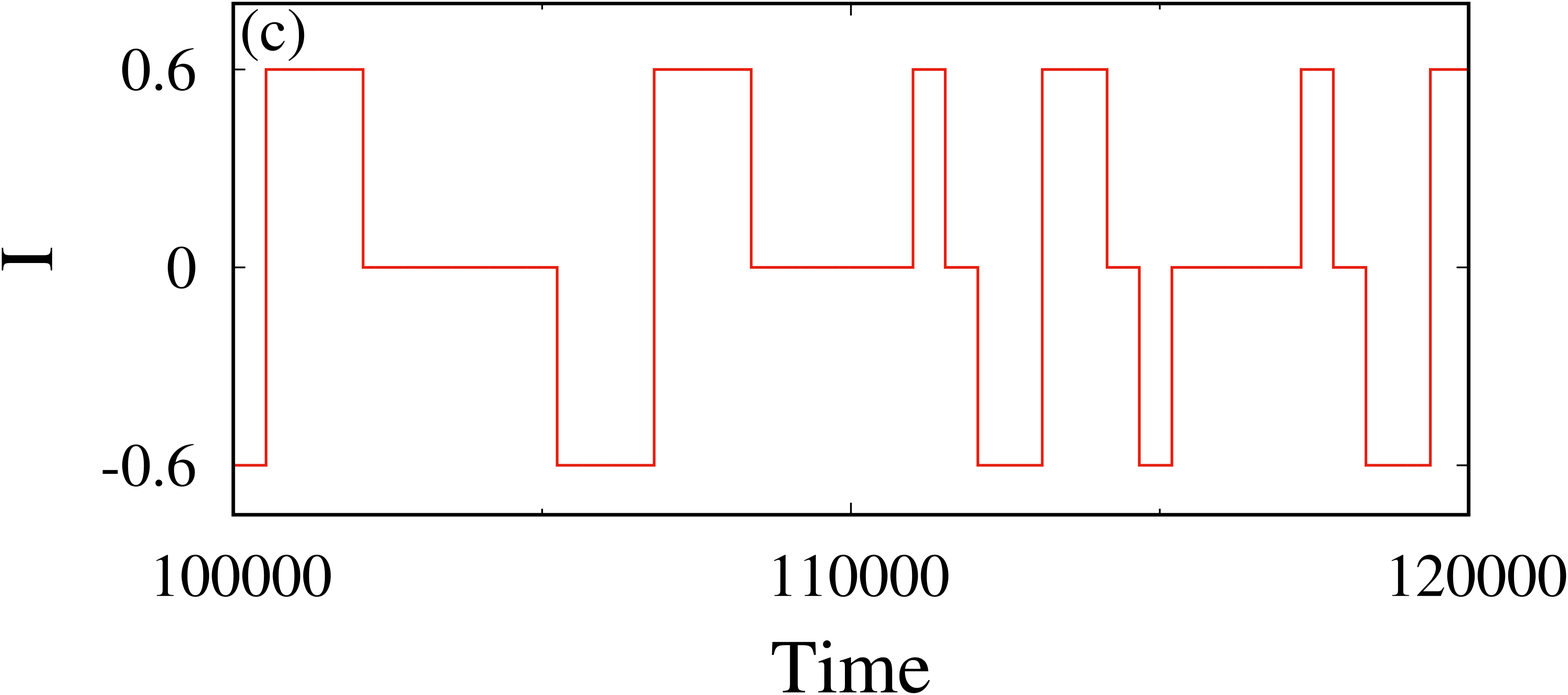}
		\caption{Panels: (a)-(b) show the two different logic inputs $I_{1}$ and $I_{2}$, respectively. (c) shows a combination of two input signals $I_{1}+I_{2}$. Input $I_{1}=I_{2}=+0.3$ when the logic input is $'1'$ and $I_{1}=I_{2}=-0.3$ when the logic input is $'0'$. The '3' level square waves with -0.6 correspond to the input set (0,0), 0 for (0,1)/(1,0) set and +0.6 for (1,1) input set.}
		\label{fig3}
	\end{figure}

	The output of the system is determined by the state $x(t)$ of system \eqref{equ1}; for example, if $\delta=0.3$ then both the inputs take the values $-0.3$ for the logical input $0$ and values $0.3$ when it is `$1$. Figs. \ref{fig3}(a) \& \ref{fig3}(b) are the two different logic input signals $ I_{1}$ \& I$_{2} $, whereas $I=I_{1}+I_{2}$ is a three-level square wave form $-0.6$ corresponding to the input set $(0,0)$, $0$ corresponding to the input sets $(0,1)/(1,0)$ and $+0.6$ corresponding to the input set $(1,1)$[see Fig.\ref{fig3}(c)].  The output is determined by the dynamical variable $x(t)$ of the system \eqref{equ1}. 	Specifically, if for $ x(t)<x^{*}$, where $ x^{*} $ is a threshold value of $x(t)$ , the response of the system is assumed to be the logical `$ 0 $' and if $ x(t)>x^{*}$, the output of the system is considered to be the logical `$ 1 $'. The value of the threshold is to be selected appropriately. In the present case the threshold value is chosen as $ x^{*}=0 $. As a result. the output of the system is considered as logical `$ 1 $', if the variable $ x(t) $  of system resides entirely positive. On the other hand for logical `$ 0 $' it resides entirely in the negative region of the phase space. 
	
	In the presence of the logic input we observe the route to logical SNA in the noise assisted periodically driven double-well Duffing oscillator. Initially we add two square waves of logic inputs in the periodically driven system and by including a moderate noise we find that the system exhibits two kinds of strange nonchaotic attractor namely i) logical SNA (which exhibits logical behavior) and ii) standard SNA (which does not exhibit logical behavior) in an optimal window regime as we point out below.

	\begin{figure}
		\centering
		\includegraphics[width=0.48\linewidth]{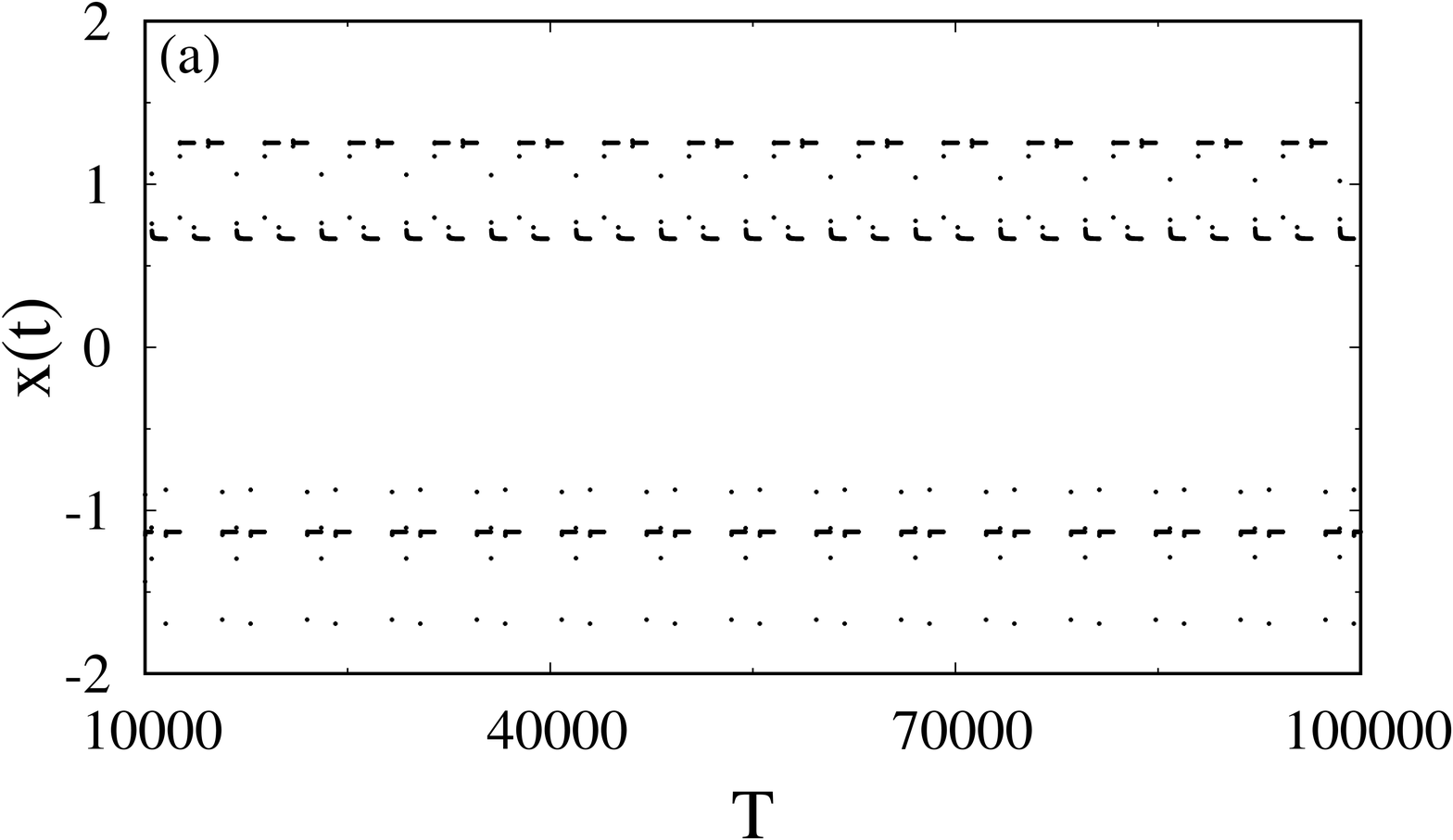}
		\includegraphics[width=0.48\linewidth]{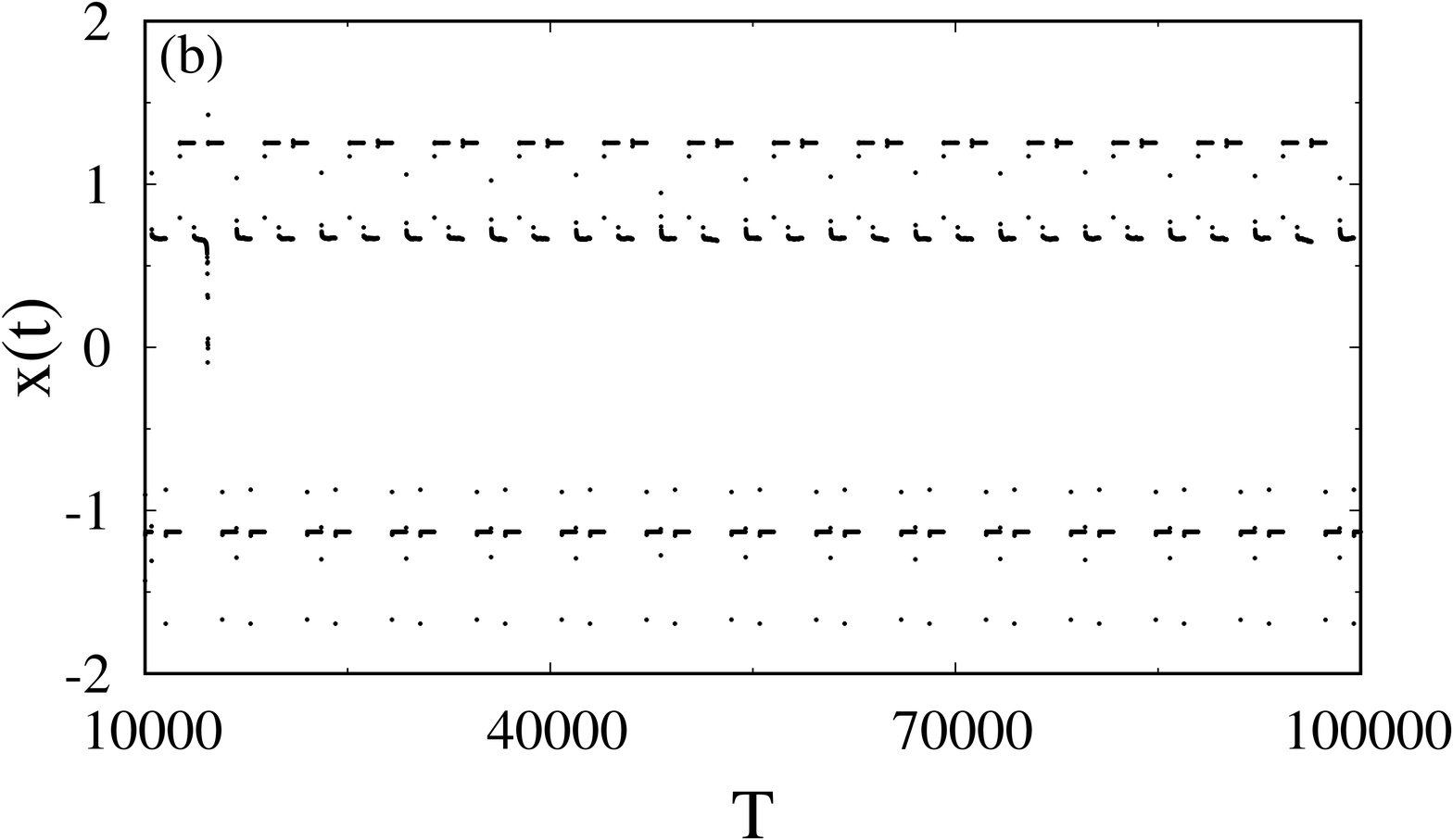}
		\includegraphics[width=0.48\linewidth]{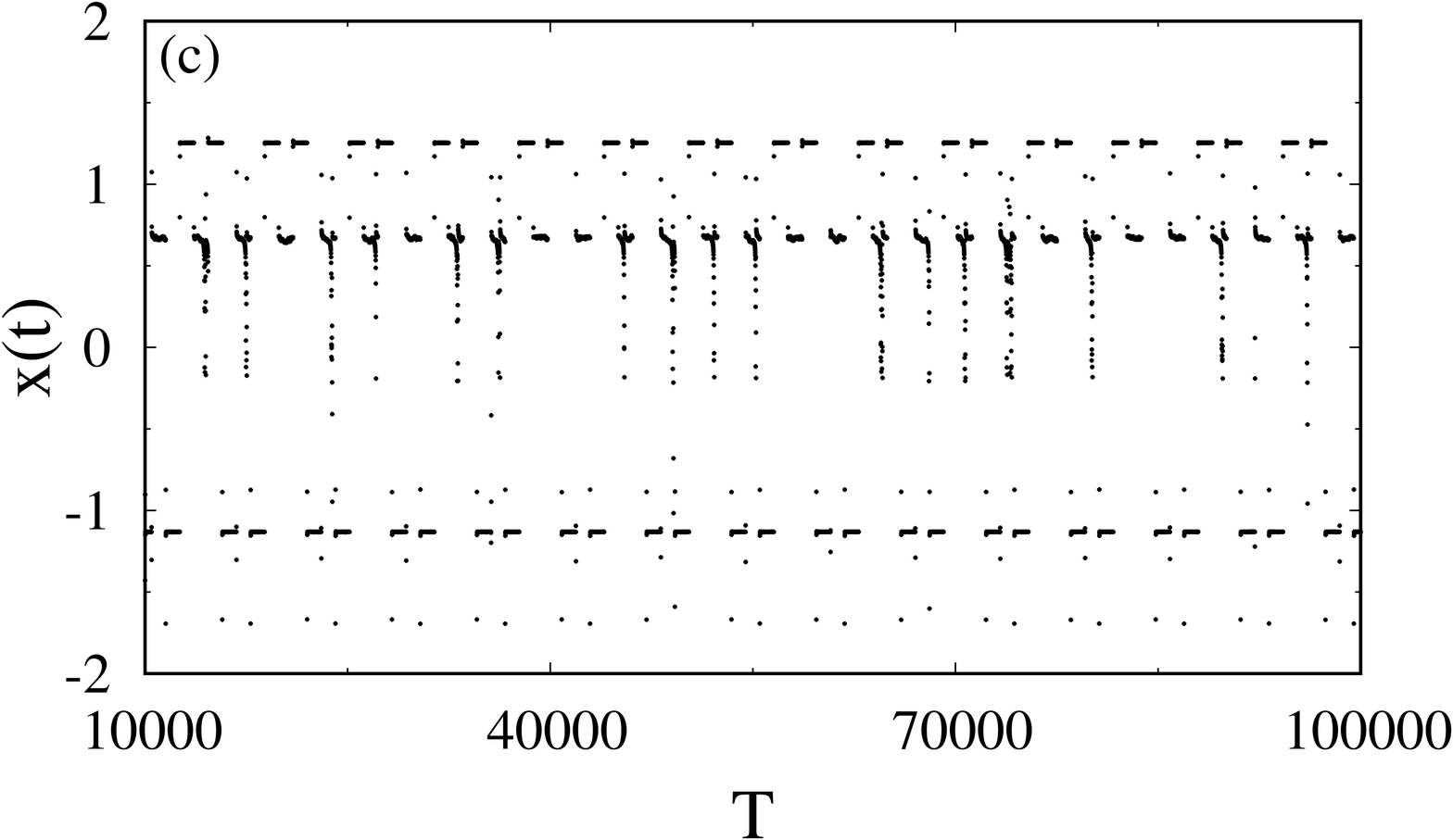}
		\includegraphics[width=0.48\linewidth]{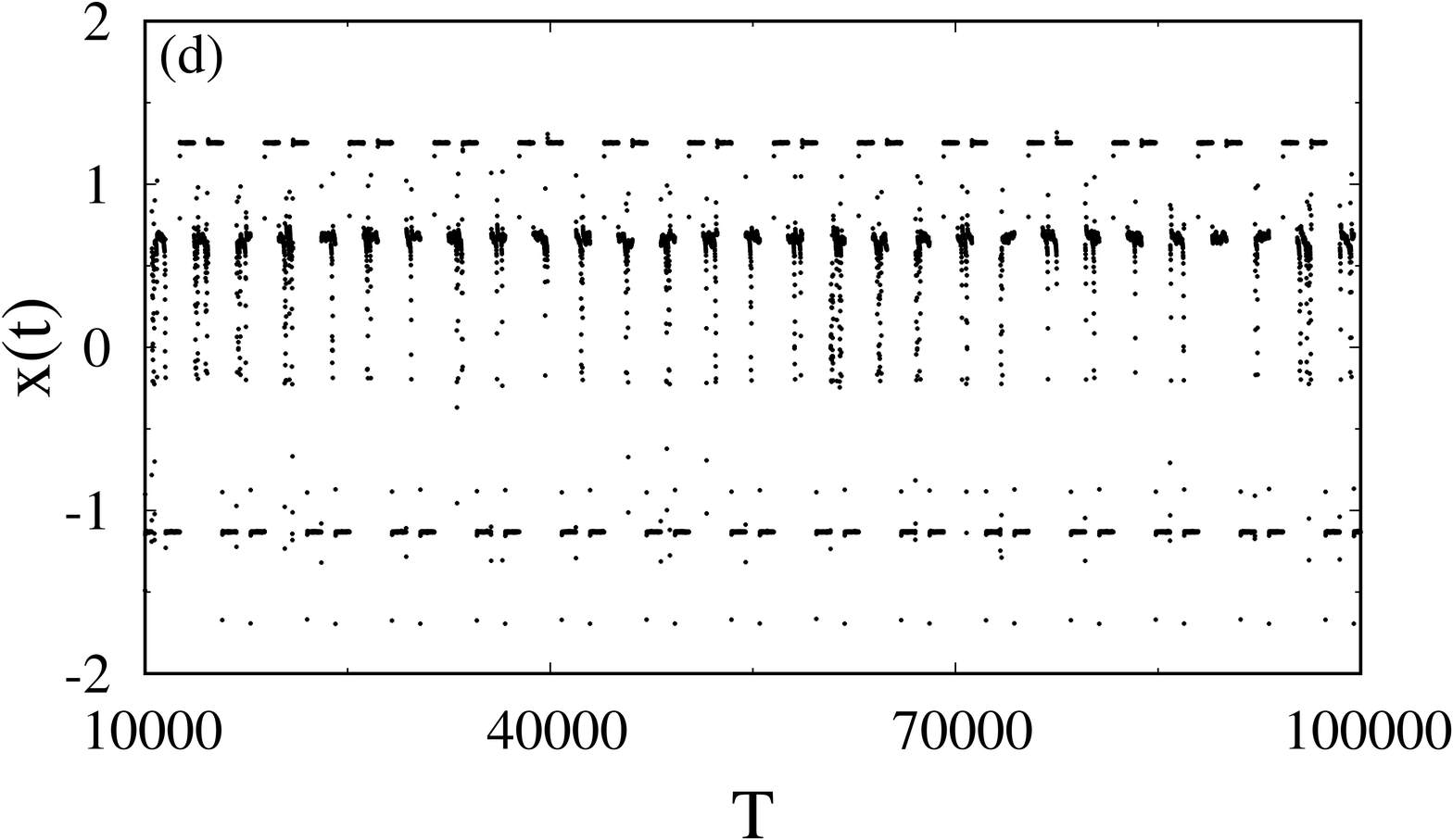}
		\caption{Panels (a)-(d) show the  Poincar\`e surface of section in the time series plane for different noise levels  $ D=0.0 $ (periodic), $ D=0.00001 $ (logical SNA), $ D=0.0001 $ (standard SNA) and $ D=0.001$ (noisy attractor) with fixed parameters $ \delta=0.3 $, $ \varepsilon=0.1 $ and $ F=0.4514 $.}
		\label{fig5}
	\end{figure}

	\begin{figure}
		\centering
		\includegraphics[width=0.8\linewidth]{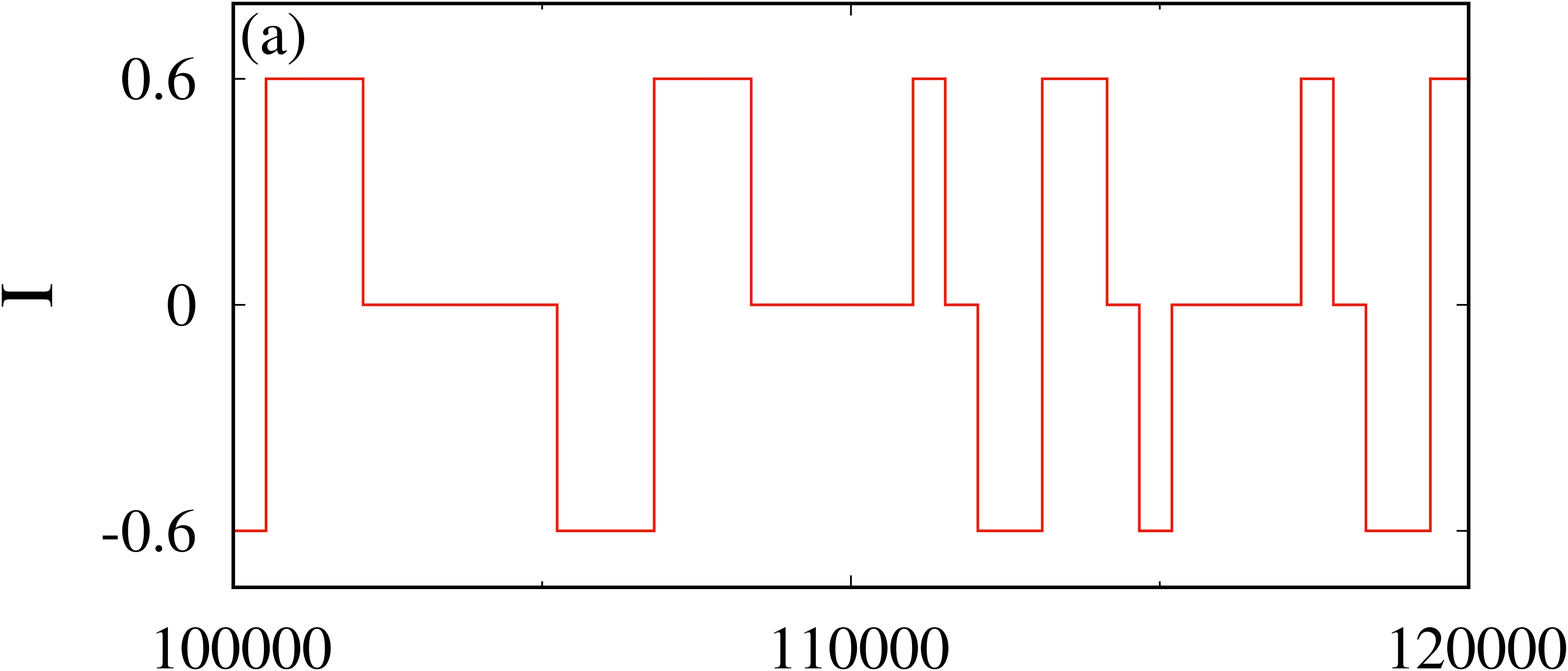}
		\includegraphics[width=0.8\linewidth]{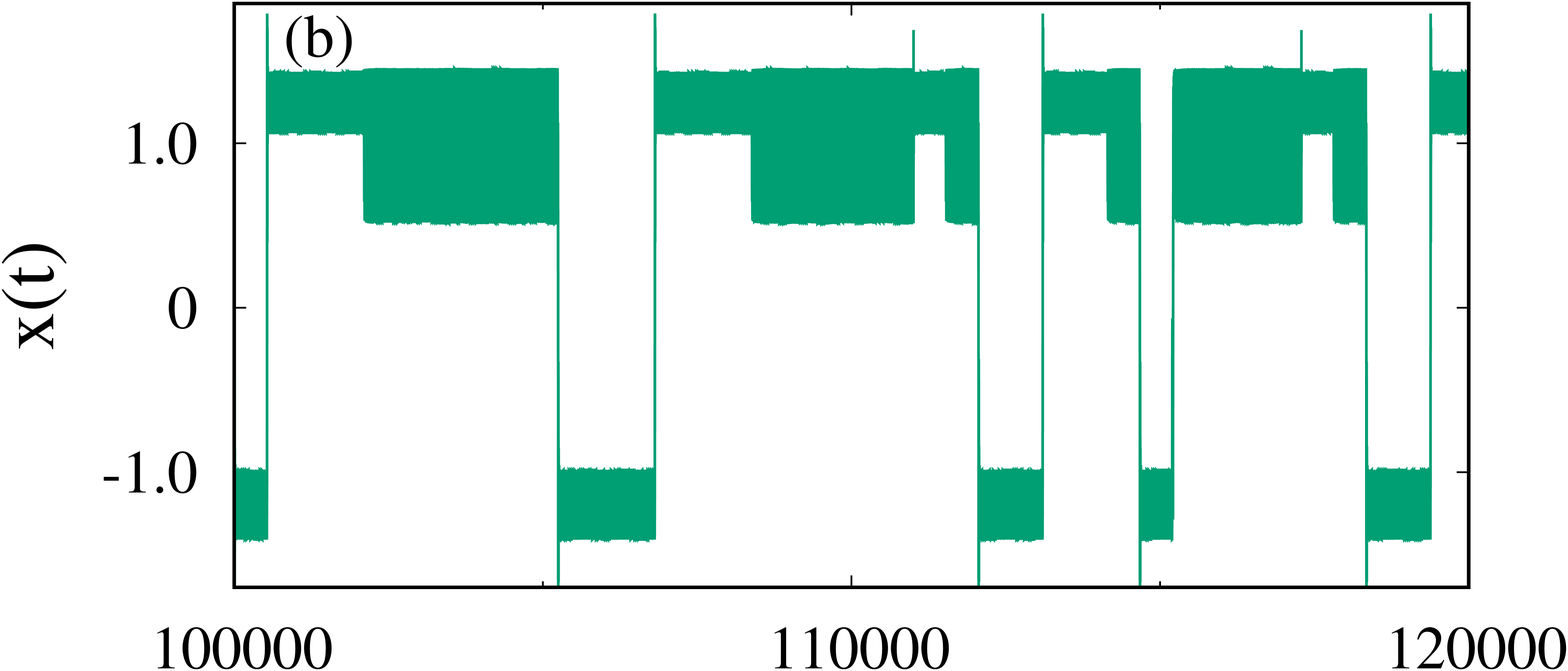}
		\includegraphics[width=0.8\linewidth]{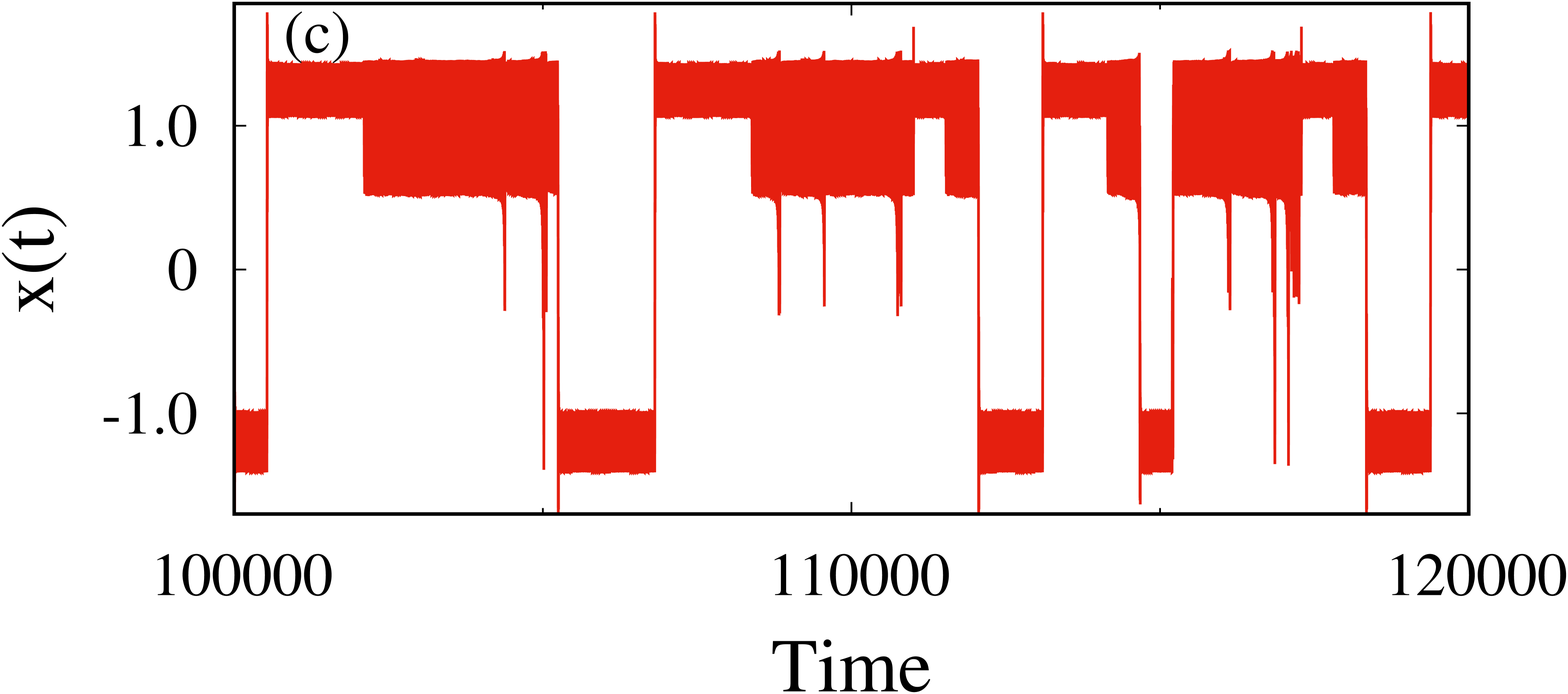}
		\caption{Panel (a) shows a combination of two input signals $I_{1}+I_{2}$. Input $I_{1}=I_{2}=+0.3$ when the logic input is $'1'$ and $I_{1}=I_{2}=-0.3$ when the logic input is $'0'$. Panels (b) \& (c) represent the corresponding dynamical response of the system x(t) under periodic forcing for different noise levels $D=0.00001$ and $D =0.0001$, respectively, with fixed bias parameter $ \varepsilon=0.1 $. }
		\label{fig4}
	\end{figure}

	\begin{figure}
	\centering
	\includegraphics[width=0.48\linewidth]{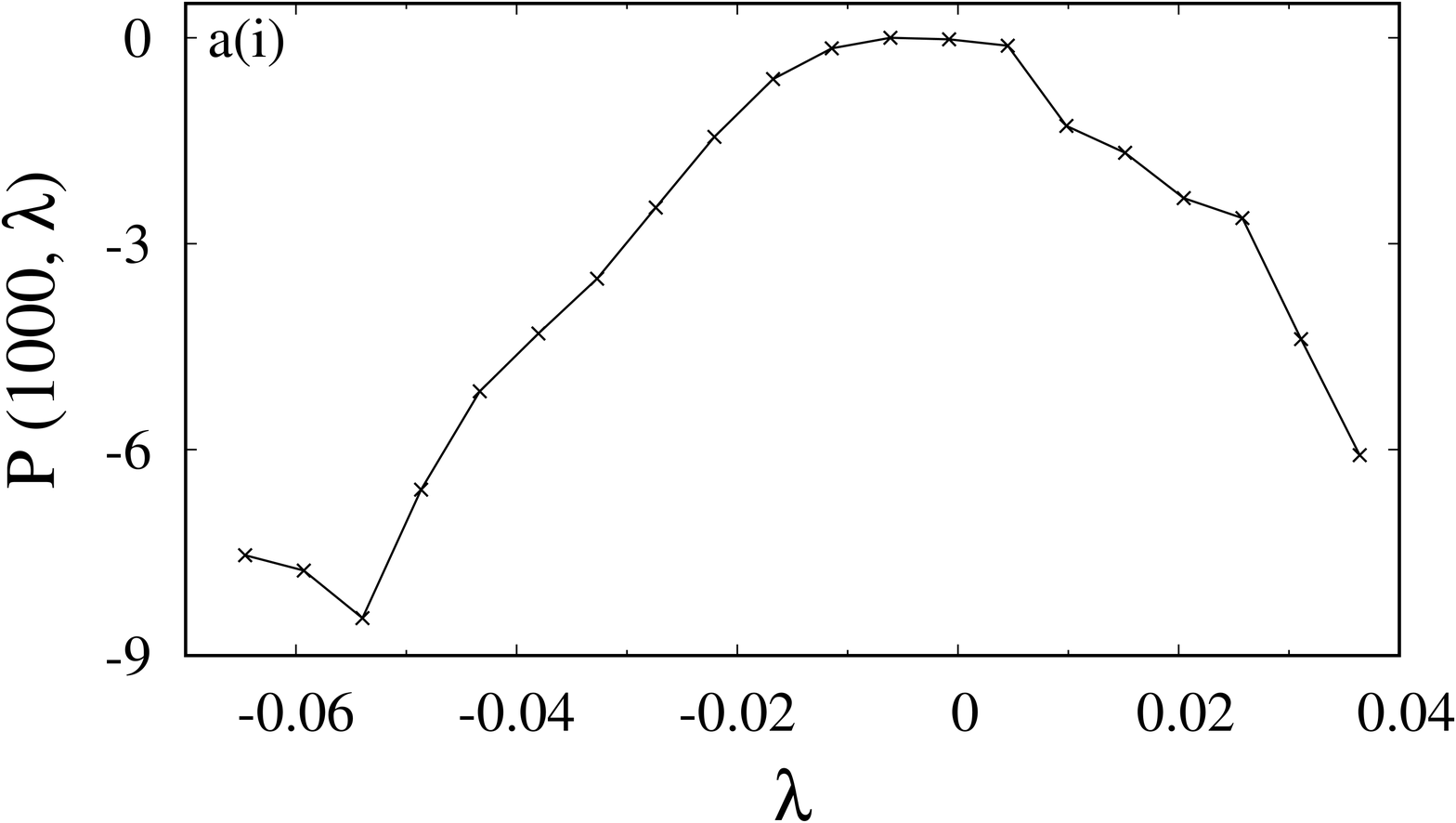}
	\includegraphics[width=0.48\linewidth]{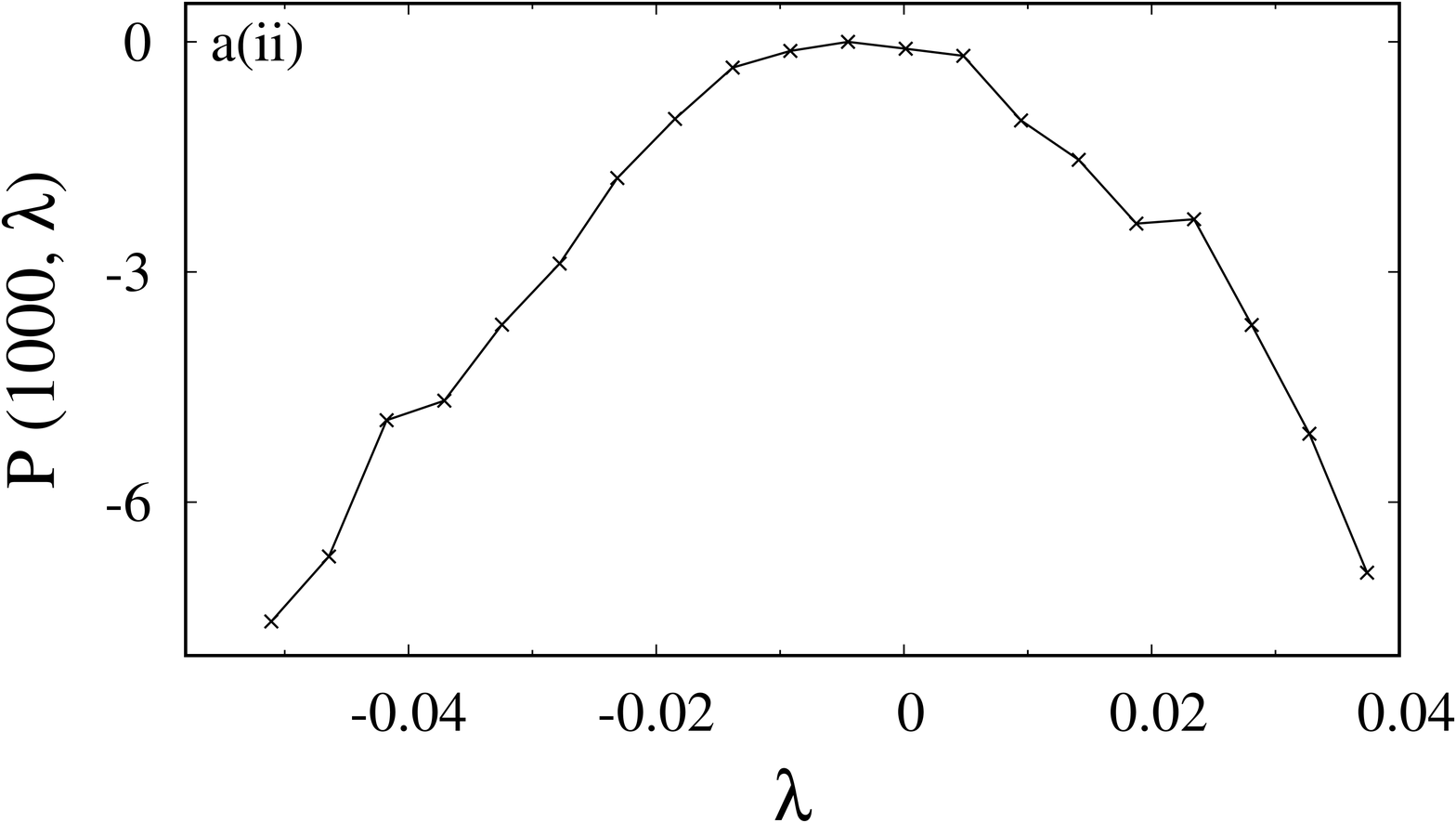}
	\includegraphics[width=0.48\linewidth]{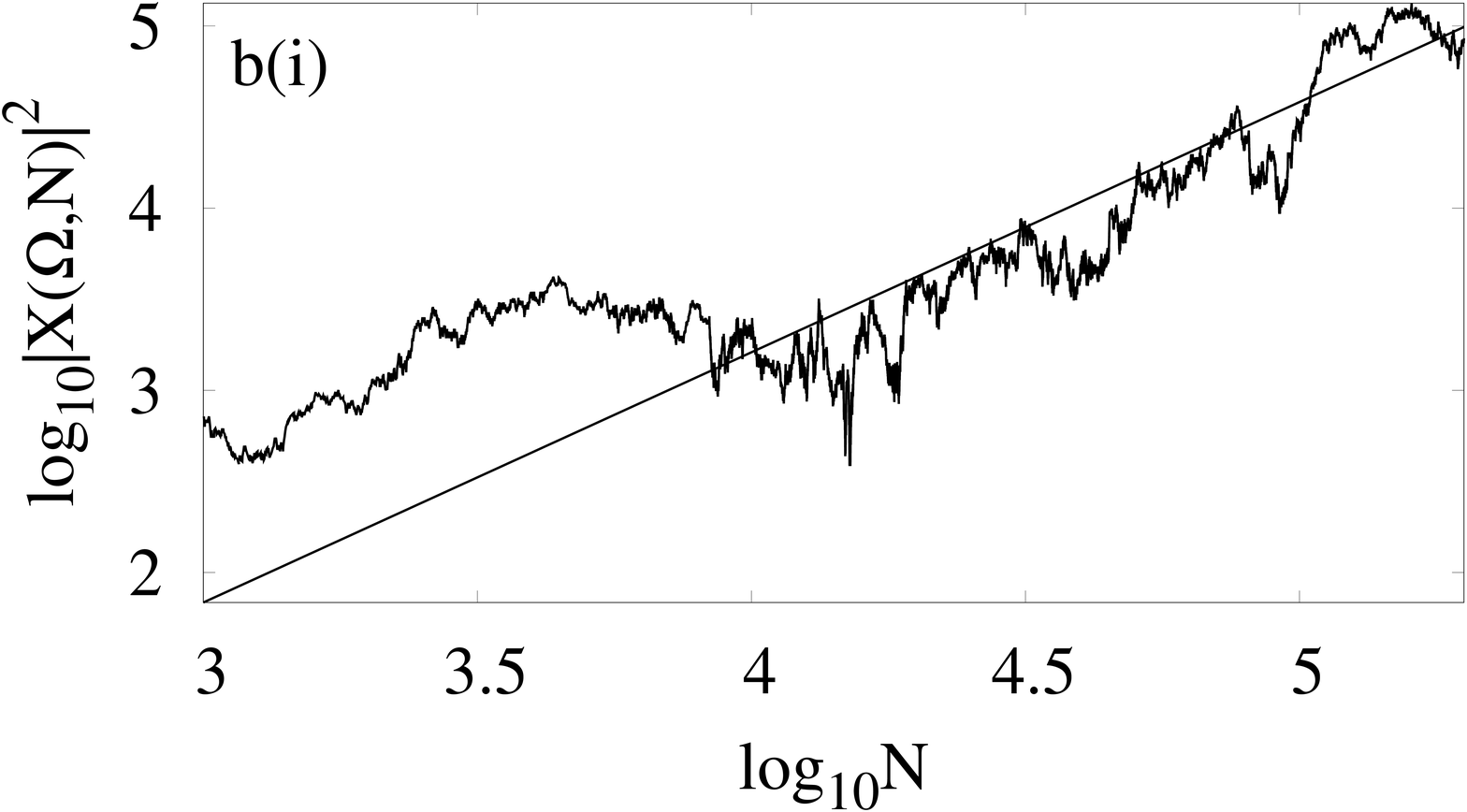}
	\includegraphics[width=0.48\linewidth]{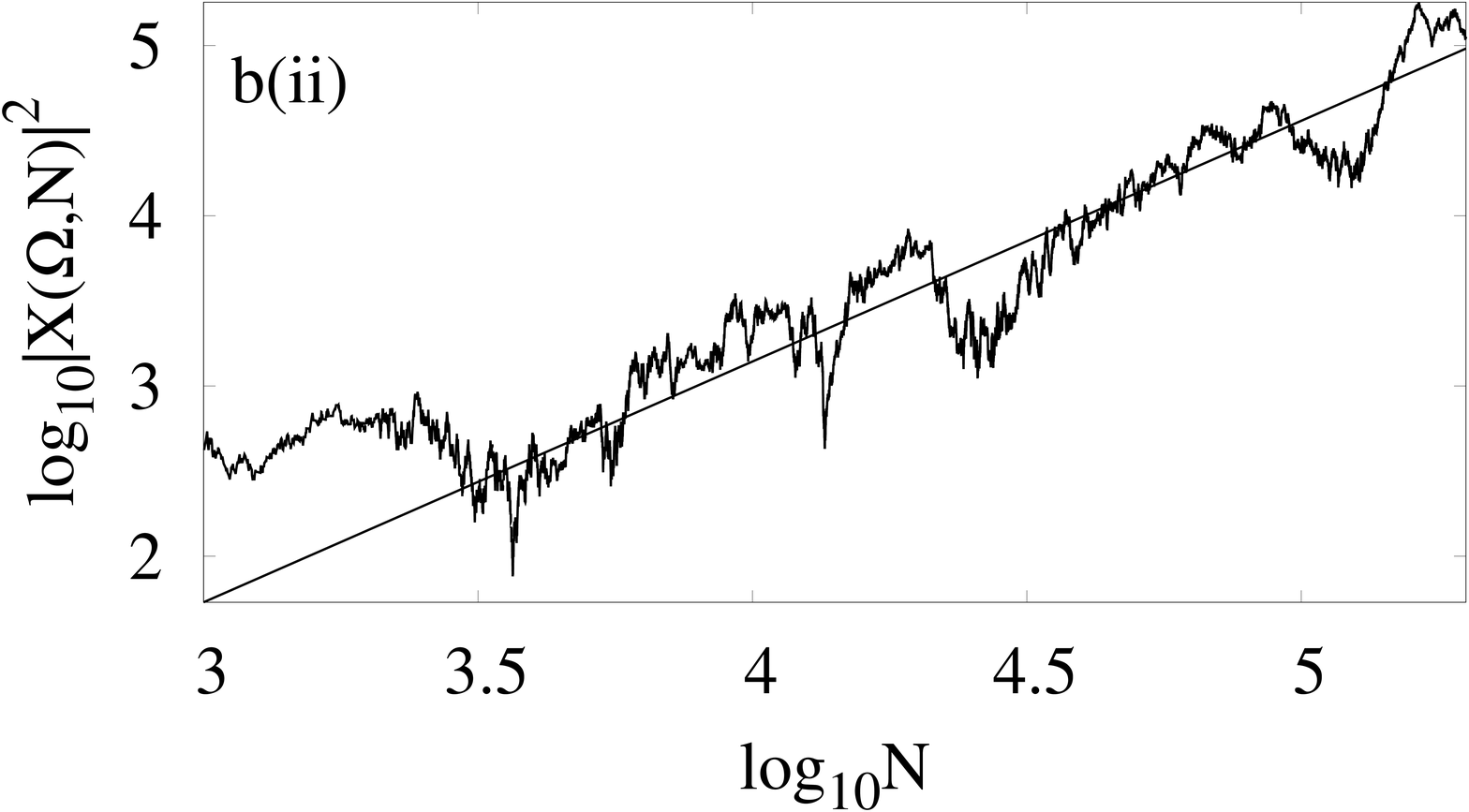}
	\includegraphics[width=0.48\linewidth]{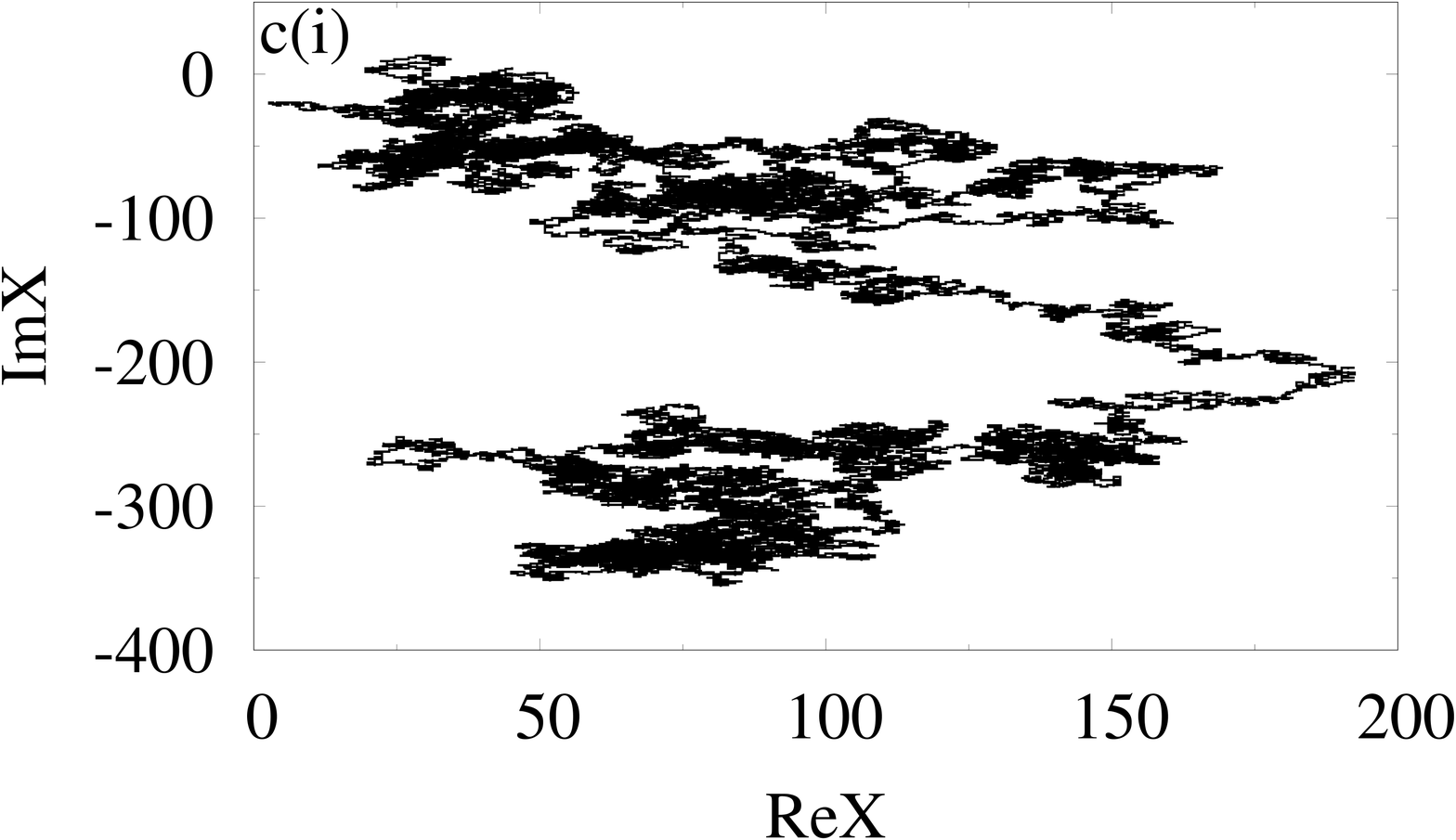}
	\includegraphics[width=0.48\linewidth]{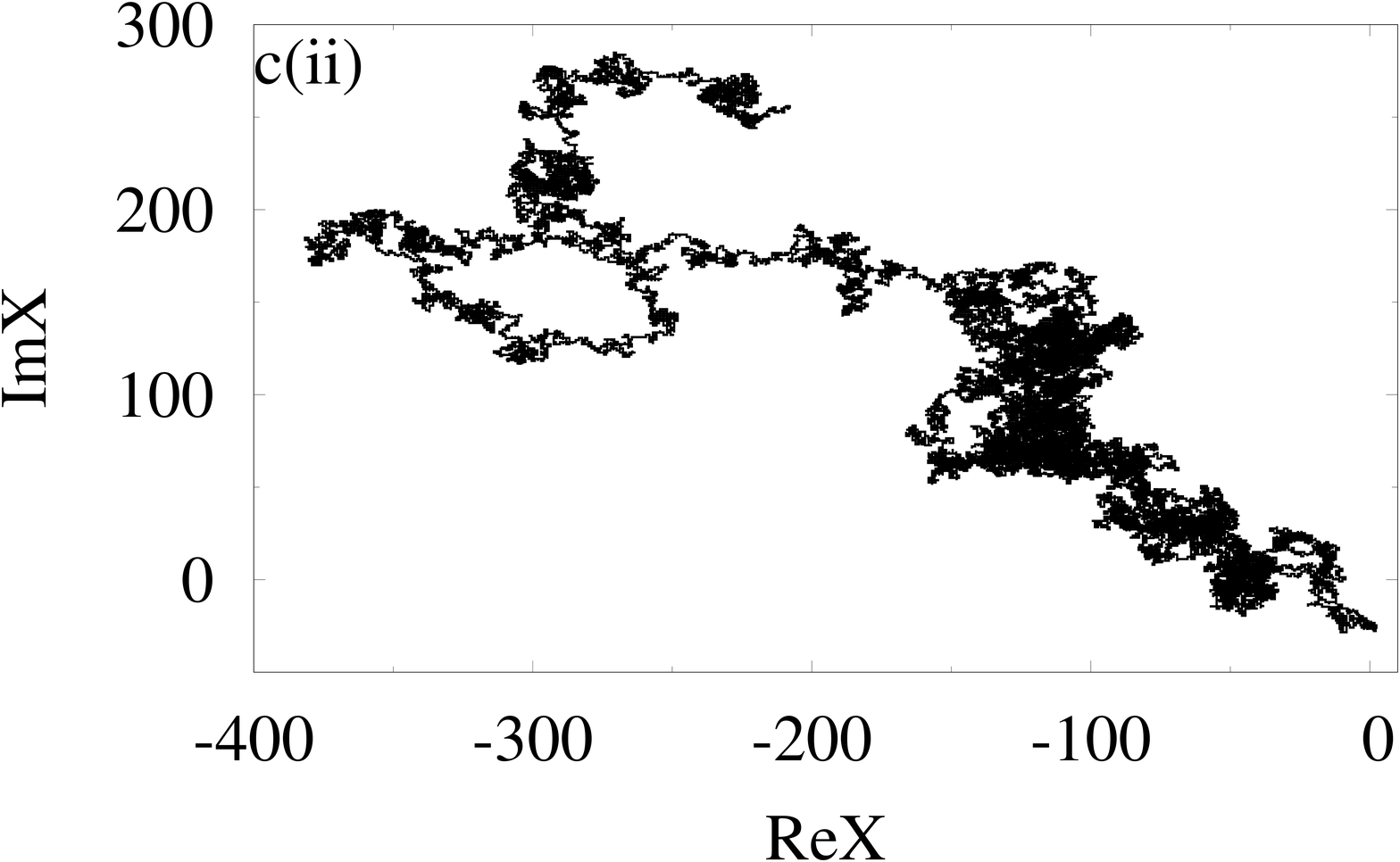}
	\caption{Projection of the logical and standard SNA attractors of Eq.\eqref{equ1}. Panels a(i) \& a(ii) show the distribution of the finite-time Lyapunov exponent in the presence of different noise strengths, $ D=0.00001 $ and $ D=0.0001 $, respectively. Panels b(i) \& b(ii) respectively show the Fourier spectrum $|X(\Omega,N)|^{2}$ vs $N^{\gamma}$ on logarithmic scale for the SNAs with noise strengths $ D=0.00001 $ with $\gamma=1.3317 $ and $ D=0.0001 $ with $\gamma=1.41441 $. Panels c(i) \& c(ii) depict fractal nature of trajectories in the complex plane corresponding to $ b(i) \& b(ii) $ for different noise strengths for fixed parameters $ F=0.4514 $, bias $ \varepsilon=0.1 $ and logic input $ \delta=0.3 $.}
	\label{fig6}
\end{figure}

	\subsection{Route to logical SNA}

	Now we fix the parameters - the strength of input signals as $ \delta=0.3 $, bias value $ \varepsilon=0.1 $, and the amplitude of the forcing as $ F=0.4514 $ and the remaining parameters at the same values as discussed in Sec.III. In the  absence of noise, the attractor is a periodic one. Since this periodic attractor lies between chaotic windows, for a small noise value the periodic attractor and the chaotic saddle are not connected dynamically. 
	
	In this case, the attractor remains a periodic one despite the inclusion of noise [see Fig.\ref{fig5}(a)]. For $ D>D_{critical}=0.000001 $, the two sets are connected dynamically. Hence, the trajectory of the system spends most of the time in the periodic regime and intermittently visits the chaotic saddle region [see Fig.\ref{fig5}(b)]. Since the chaotic saddle is part of the attractor, it is obvious that the attractor is strange and fractal. On further increase of $ D $, it is observed that the intermittent visits of the trajectory  in the chaotic saddle region increases [see Figs.\ref{fig5}(c) \& \ref{fig5}(d)]. As a result the maximal Lyapunov exponent keeps increasing from negative to positive value. The exponent still remains negative as long as  $ D<0.001 $. When we increase $ D>0.001 $, the system switches erratically between the two steady states and it slowly exhibits a noisy attractor. And on further increase of the noise strength, the attractor gets smeared out by the noise \cite{prasad1999can}. 
	
	Now, we analyze the above dynamics from the logical response point of view. When the input signal $ I=I_{1}+I_{2} $ is either (1,1) or (0,1)/(1,0) states, it is found that the attractor resides in the $ x>0 $ well and for (0,0), the attractor visits the other well $ x<0 $ as shown in Fig.\ref{fig4}(b) with the given input streams [see Fig.\ref{fig4}(a)]. If we assign the state $ x>0 $ as logical output `1' and $ x<0 $ as logical input `0', the attractor as shown in Fig.\ref{fig5}(b) and Fig.\ref{fig4}(b) is said to be the logical OR gate. On increase of $ D $ to $ D=0.0001 $, the attractor visits both the wells erratically, and thus the attractor loses its logical behavior [see Fig.\ref{fig5}(c) and Fig.\ref{fig4}(c)].
	
\subsection{Characterization of noise induced logical SNAs and standard SNAs }

To characterize further that the attractor exhibits logical behavior with SNA, we utilize the characterizations in terms of finite time Lyapunov exponents, Fourier power spectrum and fractal walks. The distributions of the FTLE with the forcing parameter $ F=0.4514 $ and noise strength $ D=0.00001 $ for logical SNA (where logic gates exist) and  $ D=0.0001 $ for standard SNA (where the gates may not show logical behavior) are shown in Figs.\ref{fig6}a(i) and \ref{fig6}a(ii), respectively. The distribution of FTLE shows that it is present mostly in the negative region for both the logical and standard SNAs but more so for the logical SNA. It is evident from the spectral distribution for the logical and standard SNAs, the power spectrum obeys the relation as $|X(\Omega,N)|^{2} \sim N^{\gamma}$,  where $\gamma=1.3317$ for the logical SNA [see Fig.\ref{fig6}b(i)] and $\gamma=1.41441$ for the standard SNA [see Fig.\ref{fig6}b(ii)], respectively. Further the fractal walk of the trajectories in the complex \emph{(ReX, ImX)} plane are demonstrated for logical and standard SNAs [see Figs.\ref{fig6}c(i) and \ref{fig6}c(ii)]. All the characterizations clearly indicate the noise-induced logical and standard SNAs for the Duffing oscillator. 

\begin{figure}[h!]
	\centering
	\includegraphics[width=0.48\linewidth]{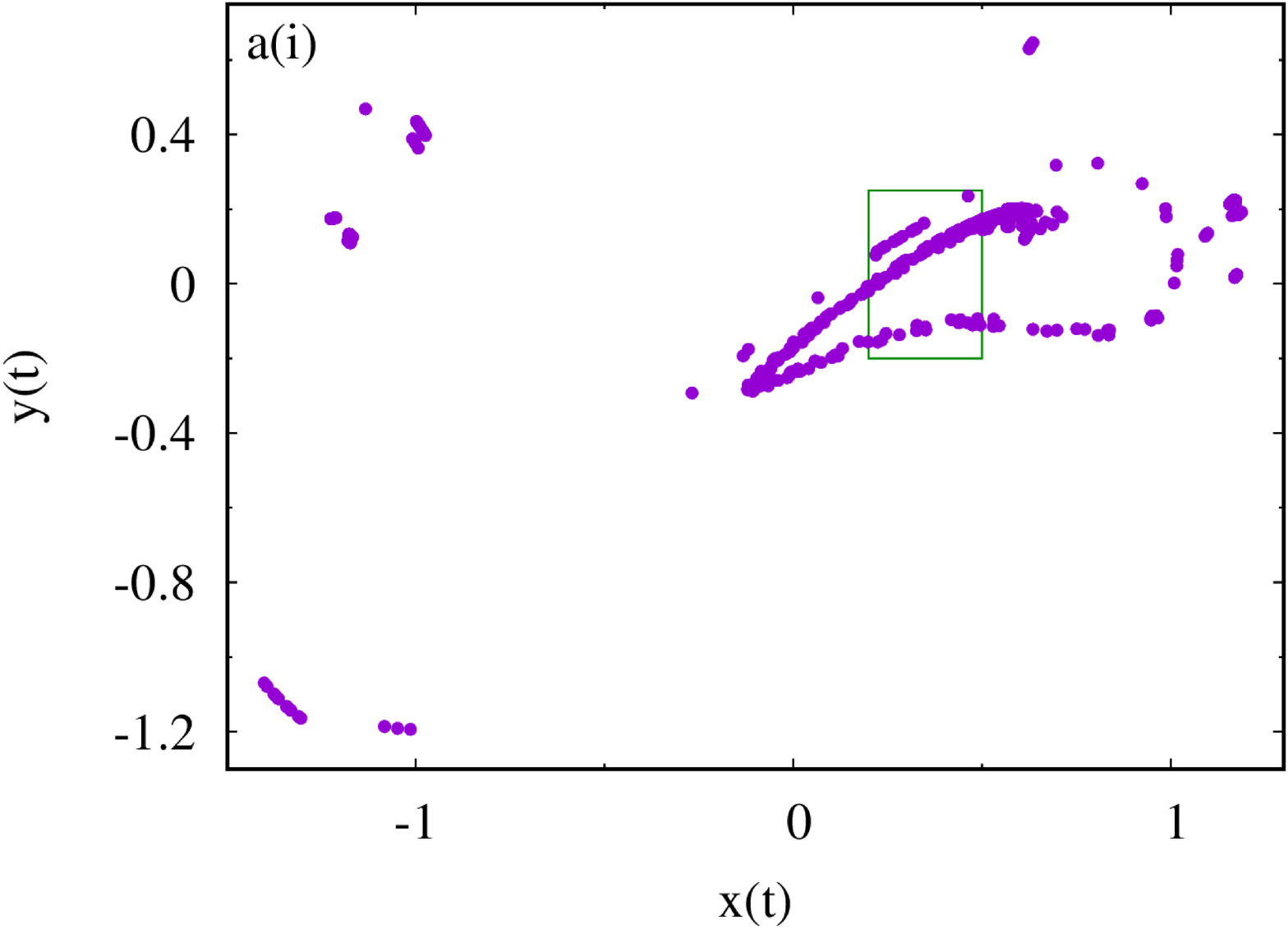}
	\includegraphics[width=0.48\linewidth]{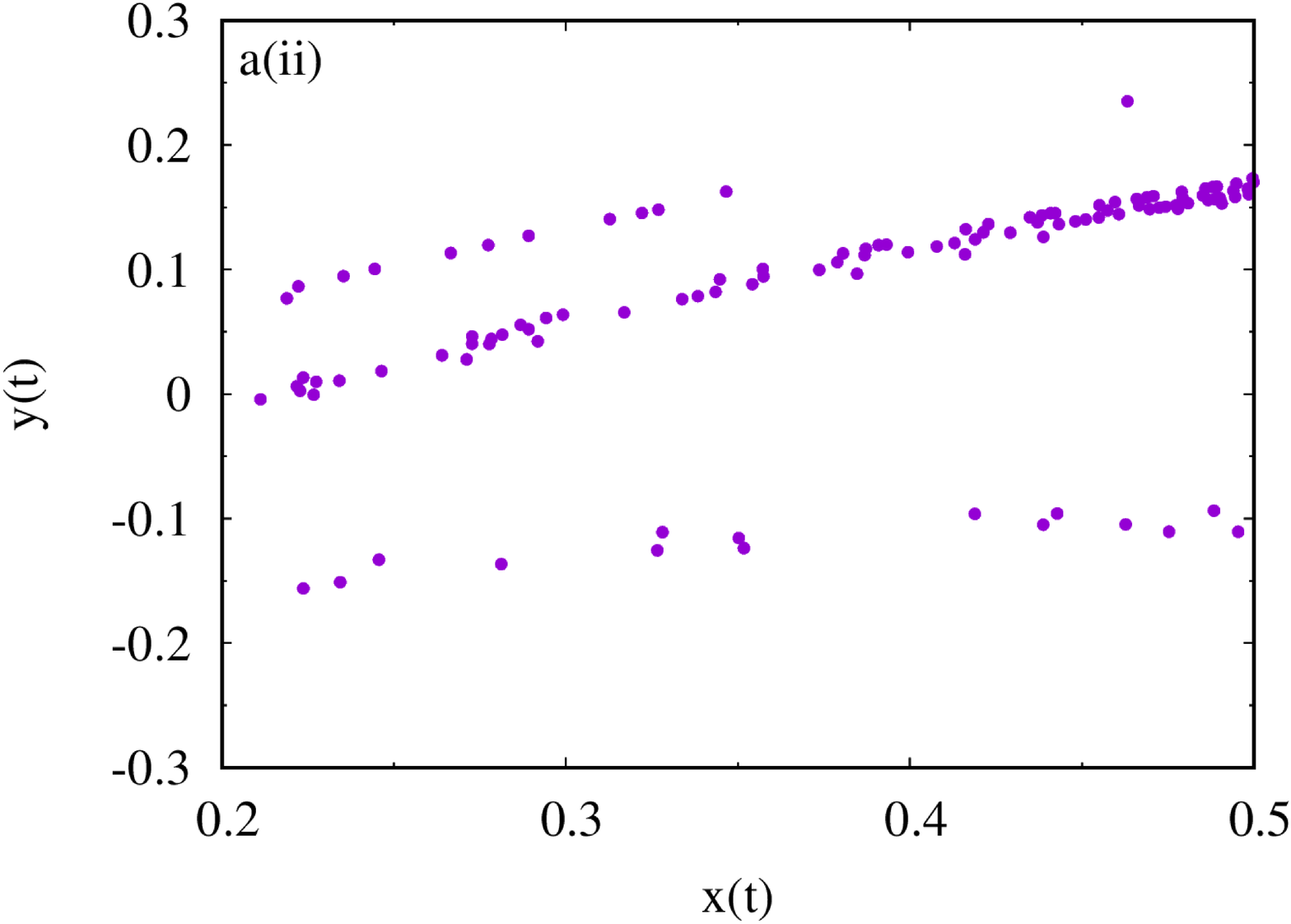}
	\includegraphics[width=0.48\linewidth]{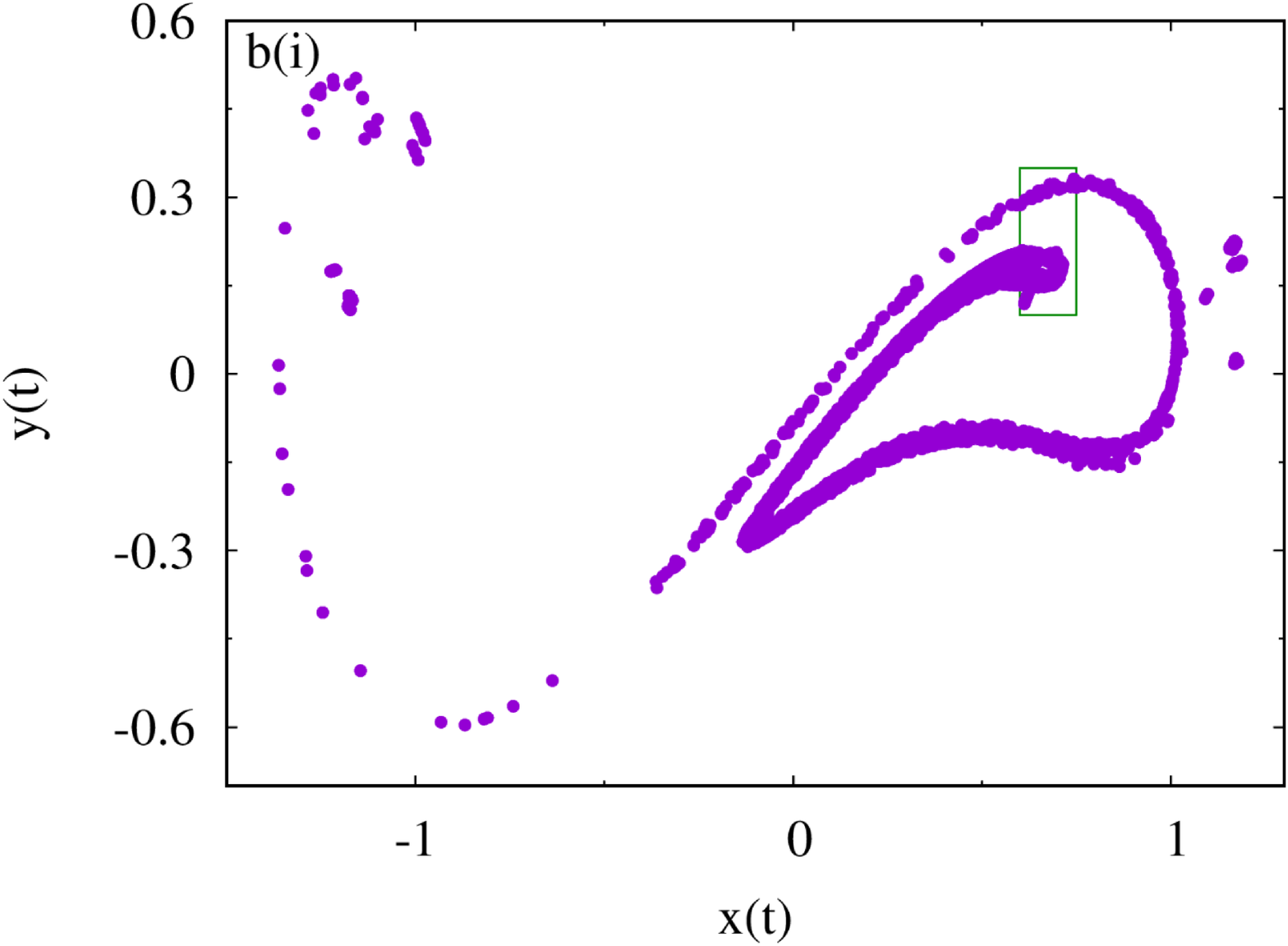}
	\includegraphics[width=0.48\linewidth]{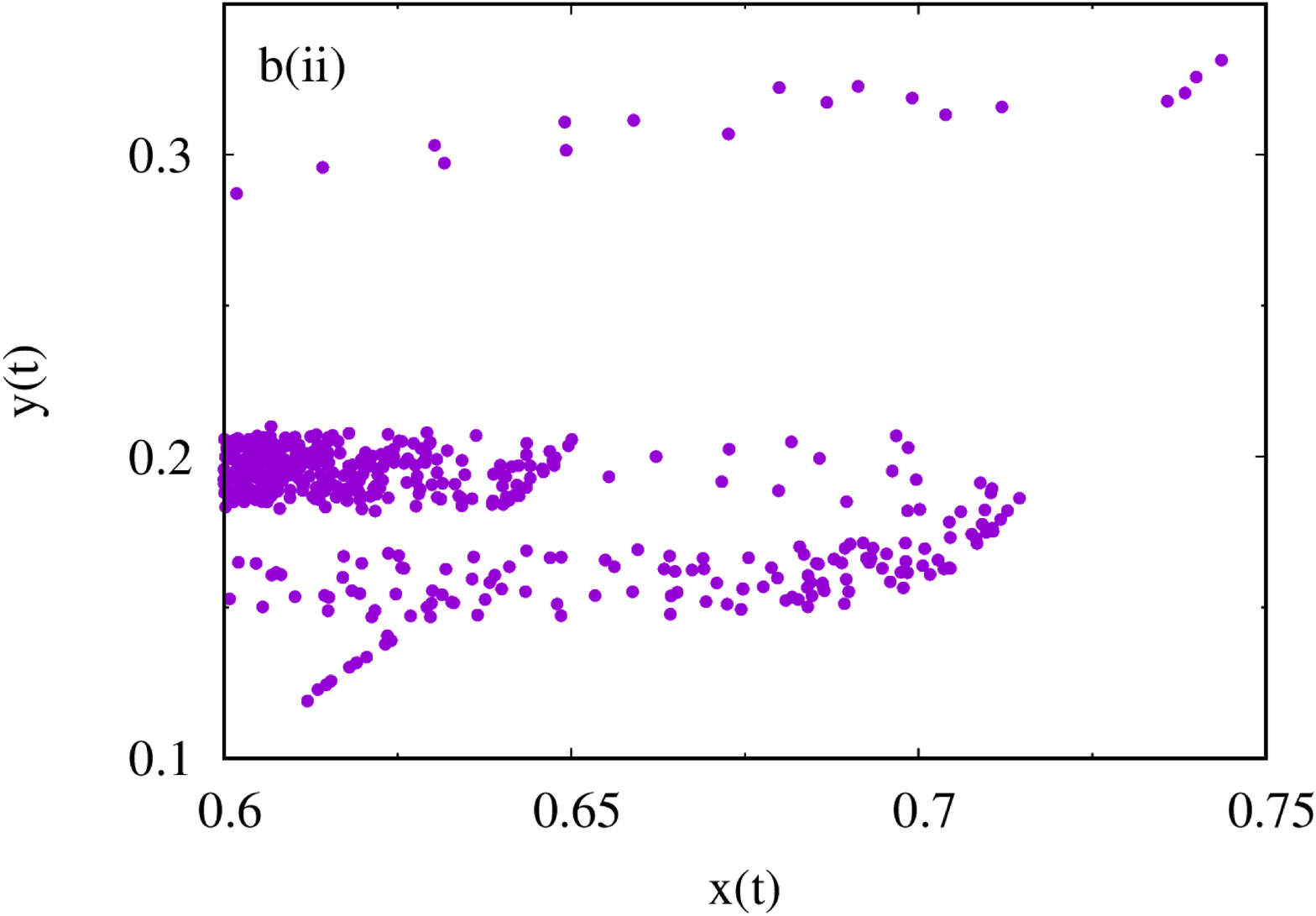}
	\includegraphics[width=0.48\linewidth]{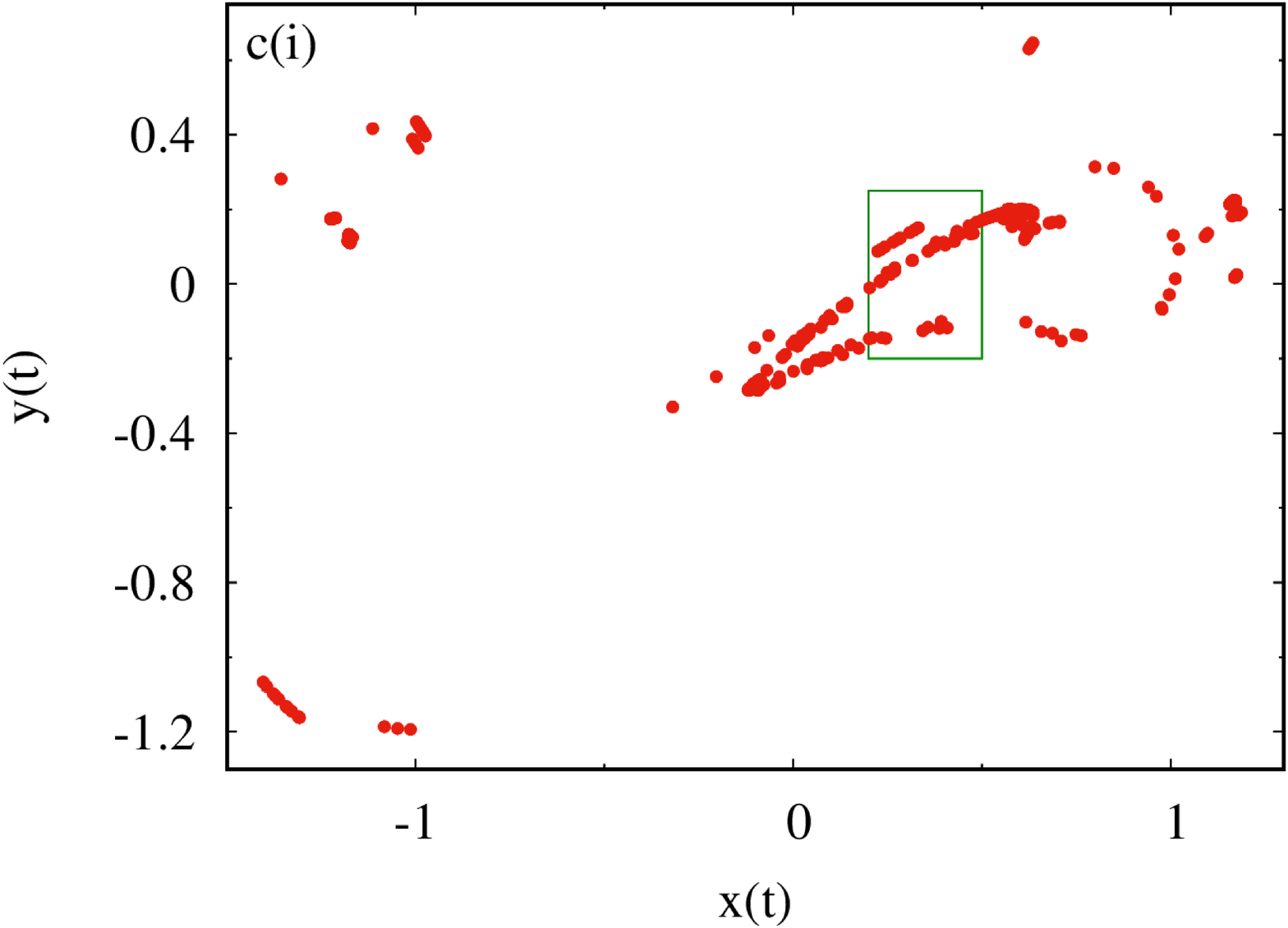}
	\includegraphics[width=0.48\linewidth]{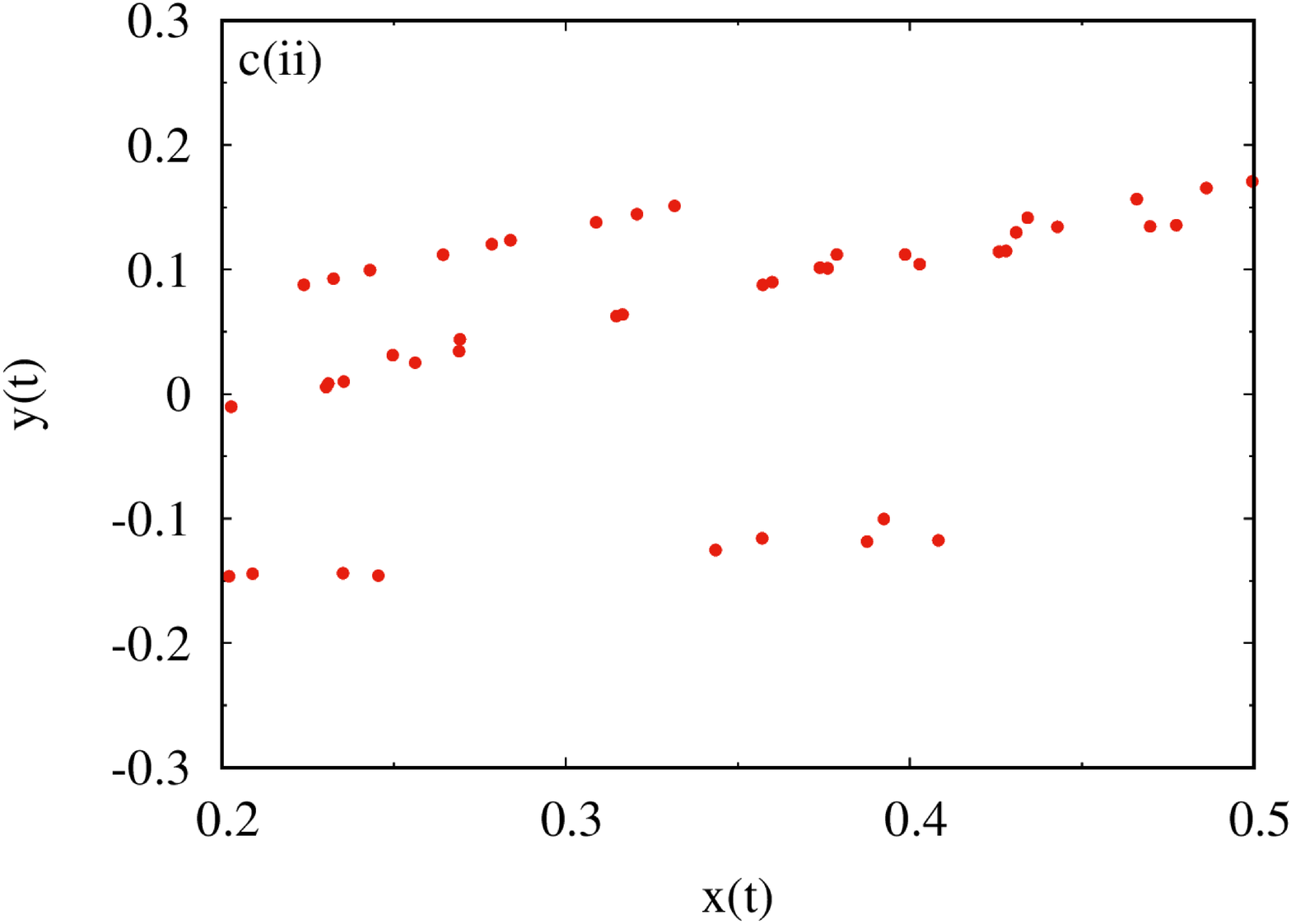}
	\includegraphics[width=0.48\linewidth]{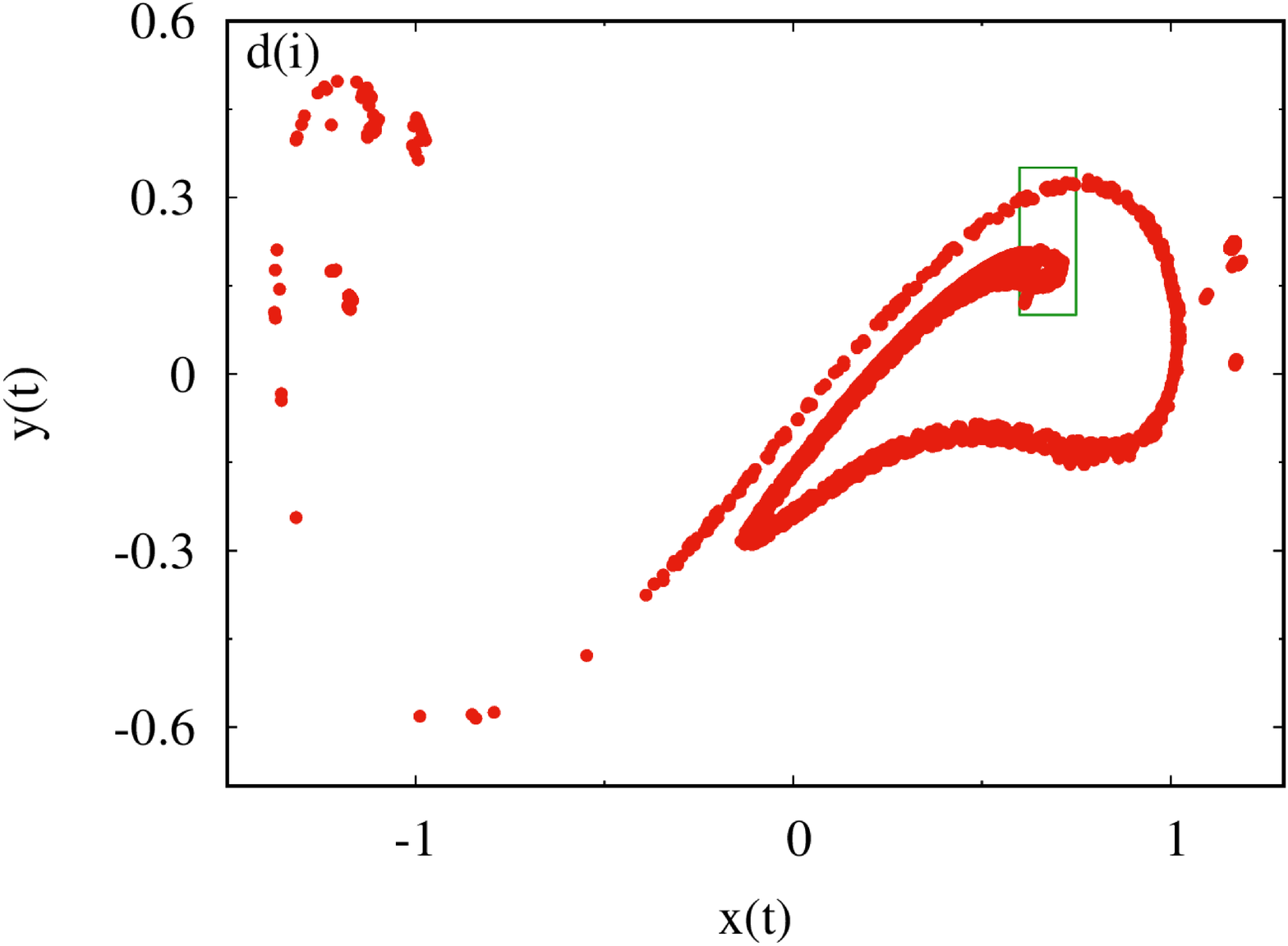}
	\includegraphics[width=0.48\linewidth]{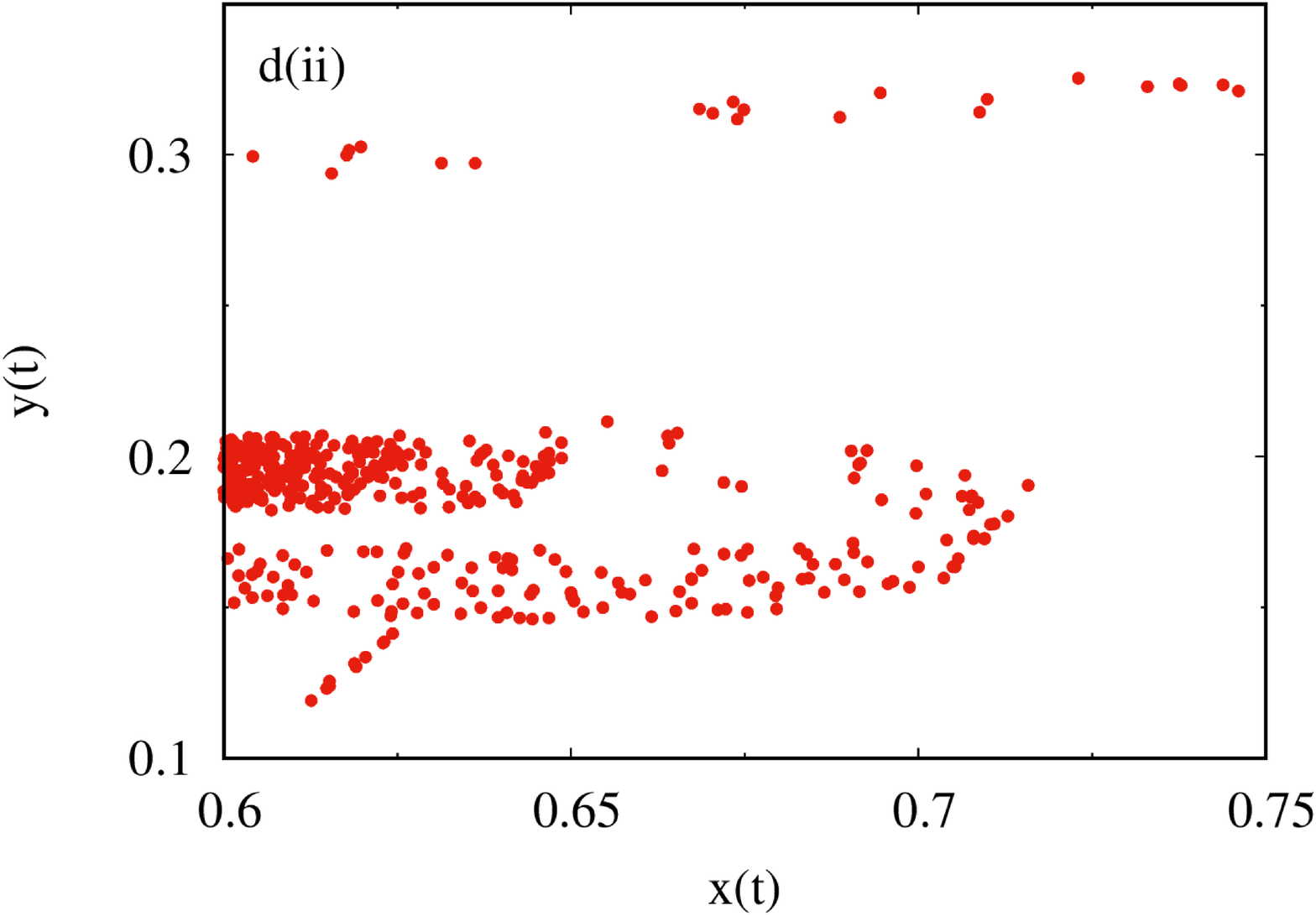}
	\caption{Poincar\'e section of the phase space  with bias $ \varepsilon=0.1 $ and logic pulse $ \delta =0.3 $ for different white noise signals of amplitude $ D=0.00001 $ (logical SNA) and $ D=0.0001 $ (standard SNA) :-  Panels a(i) \& b(i) represent a single trajectory of logical and standard SNA, respectively, and Panels c(i) \& d(i) represent the snapshot attractors from 10,000 trajectories of logical and standard SNAs, respectively.  Panels a(ii), b(ii), c(ii) and d(ii) correspond respectively to the blow-up parts of single and snapshot attractors from logical and standard SNA.}
	\label{fig12}
\end{figure}

Noise-induced SNAs and fractal snapshot attractors can also occur due to the effect of logic signals in the periodically driven Duffing oscillator system.  We have again constructed the snapshot attractor as discussed in Sec.III for this case too. It is demonstrated clearly in Figs.\ref{fig12}(a) and (b) of a single trajectory of the logic SNA and standard SNA respectively, do not reveal the fractal structure  while the snapshot attractors [see Figs.\ref{fig12}(c) and (d)] are apparently fractal.

\subsection{Probability of obtaining logic gates}

\begin{figure}[h!]
	\centering
	\includegraphics[width=0.8\linewidth]{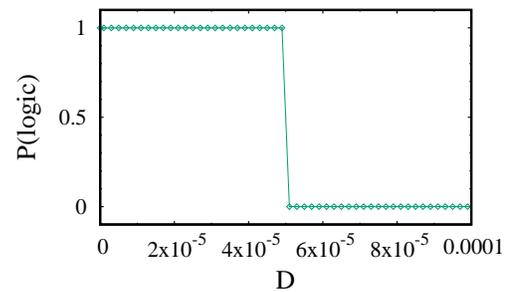}
	\caption{Probability of obtaining logic output for various noise strengths $ D $ with fixed forcing parameter $  F=0.4514 $, logic input $ \delta=0.3 $ and $ \varepsilon=0.1 $.}
	\label{fig10}
\end{figure}

The consistency of obtaining logic gates can be confirmed by estimating the probability of the desired output for different noise strengths, where P(Logic) is estimated by calculating the ratio of the number of runs which gives the correct logic output to the total number of runs with different input streams. When the probability P(logic) is 1, the system reproduces completely reliable logic gates. This notion of probability would help to identify which region will explicitly show the logic operation for different noise strengths $ 0<D<0.0001 $.  It is obvious from Fig.\ref{fig10}, that for an optimal window of noise intensity one can get consistently the logical responses as output in the system.

\begin{figure}[h!]
	\centering
	\includegraphics[width=0.48\linewidth]{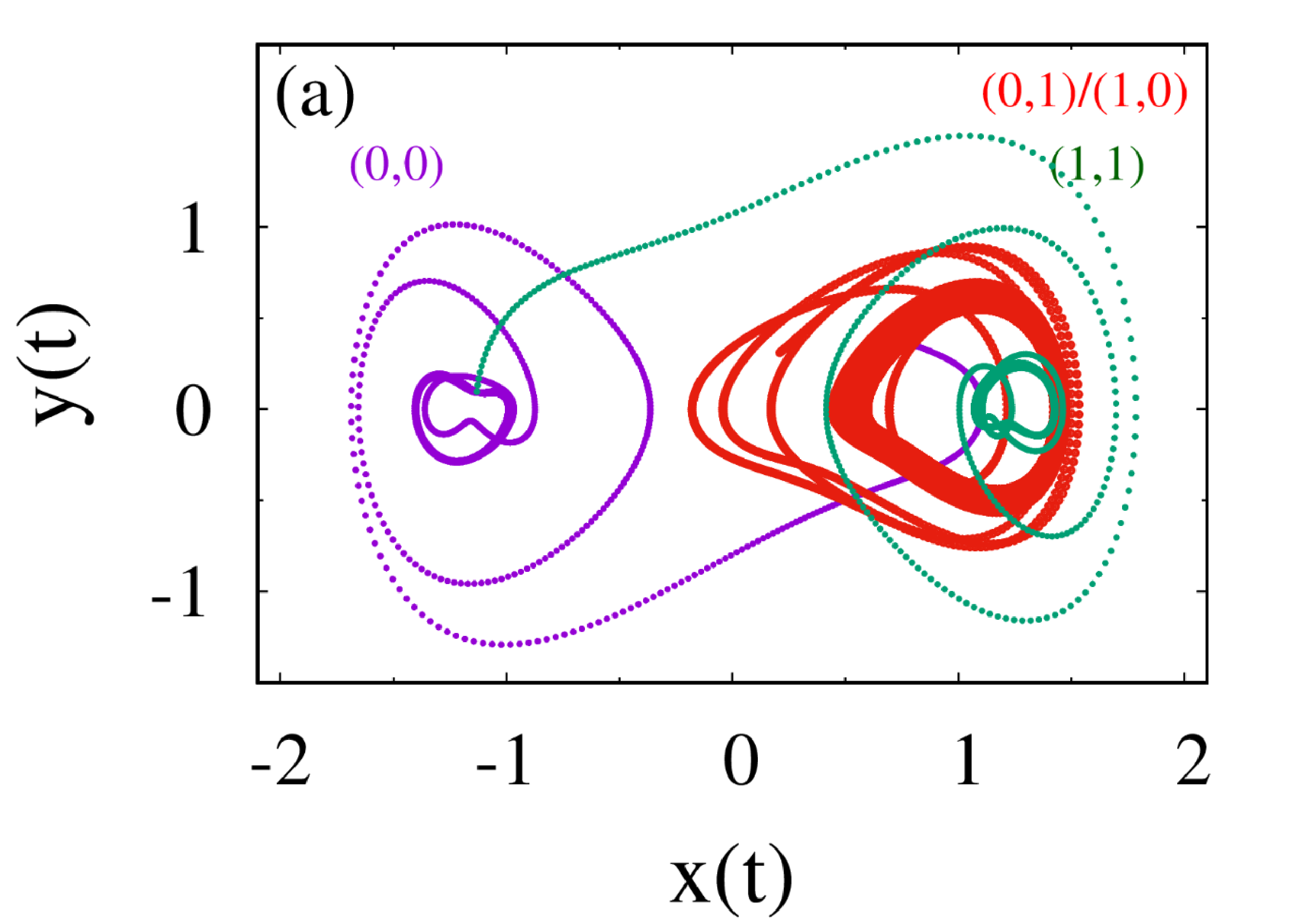}
	\includegraphics[width=0.48\linewidth]{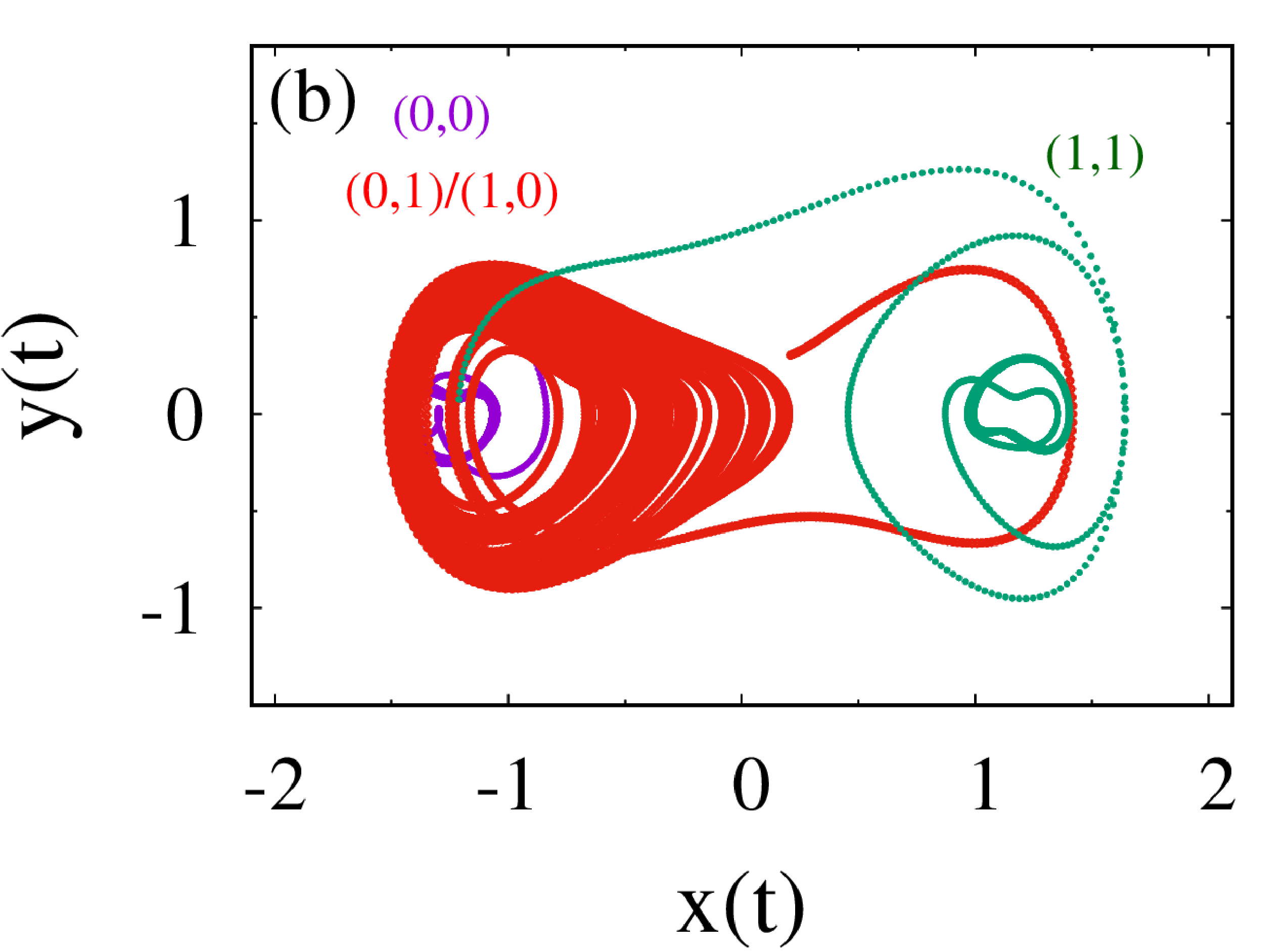}
	\caption{Phase space plane for various values of `$\varepsilon$'. Panels (a) and (b)  represent the logical OR gate with $\varepsilon=0.1$ and  logical AND gate with $\varepsilon=-0.1$, respectively, for fixed parameters $F=0.44$, $ D=0.00001 $ and $ \delta=0.3$.}
	\label{fig11}
\end{figure}

\begin{figure}
	\centering
	\includegraphics[width=0.8\linewidth]{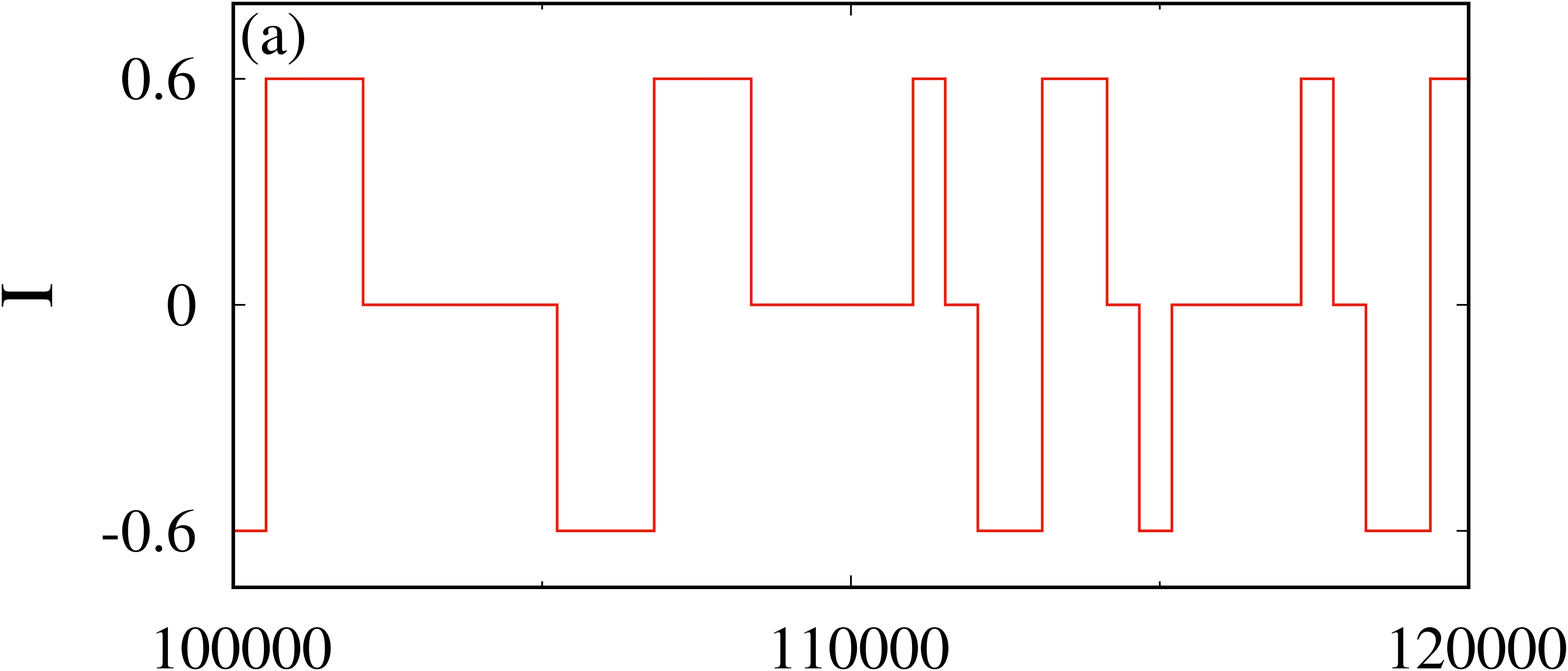}
	\includegraphics[width=0.8\linewidth]{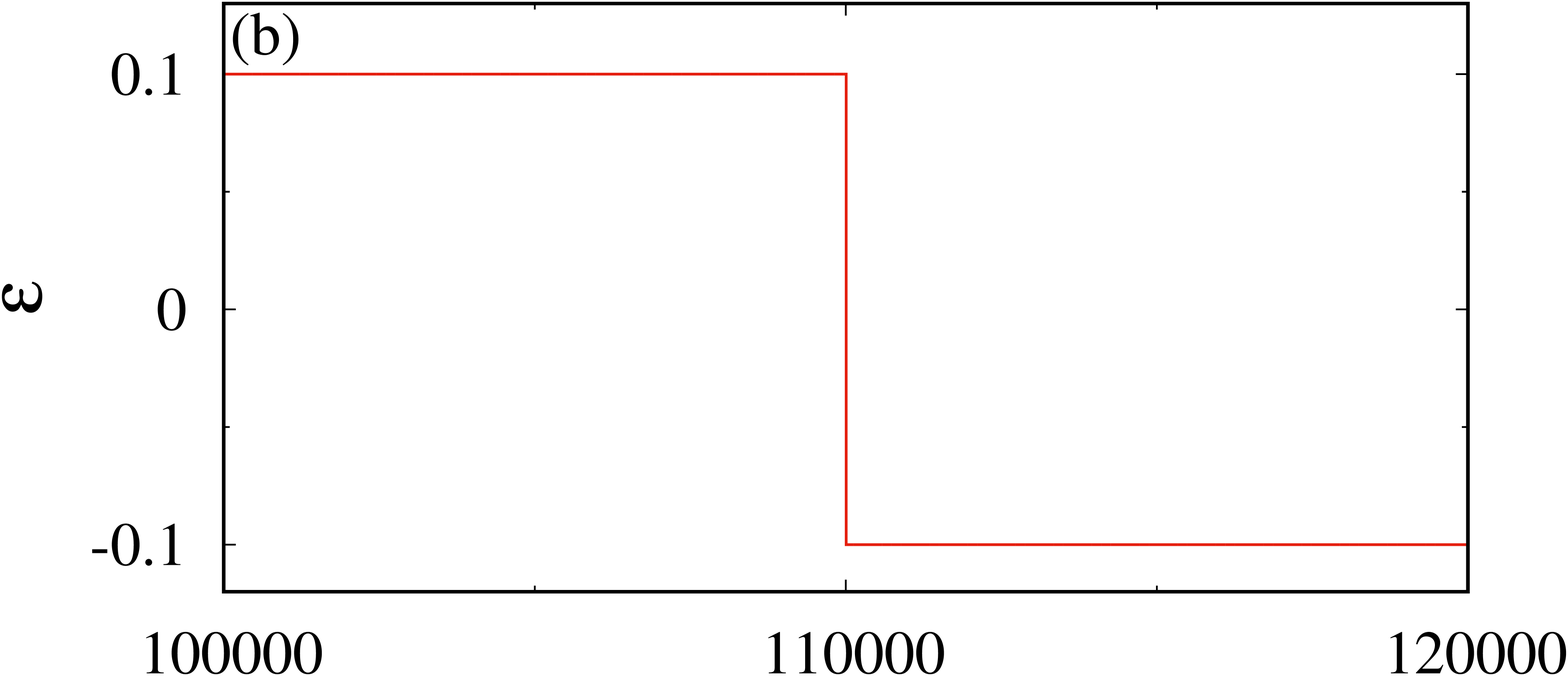}
	\includegraphics[width=0.8\linewidth]{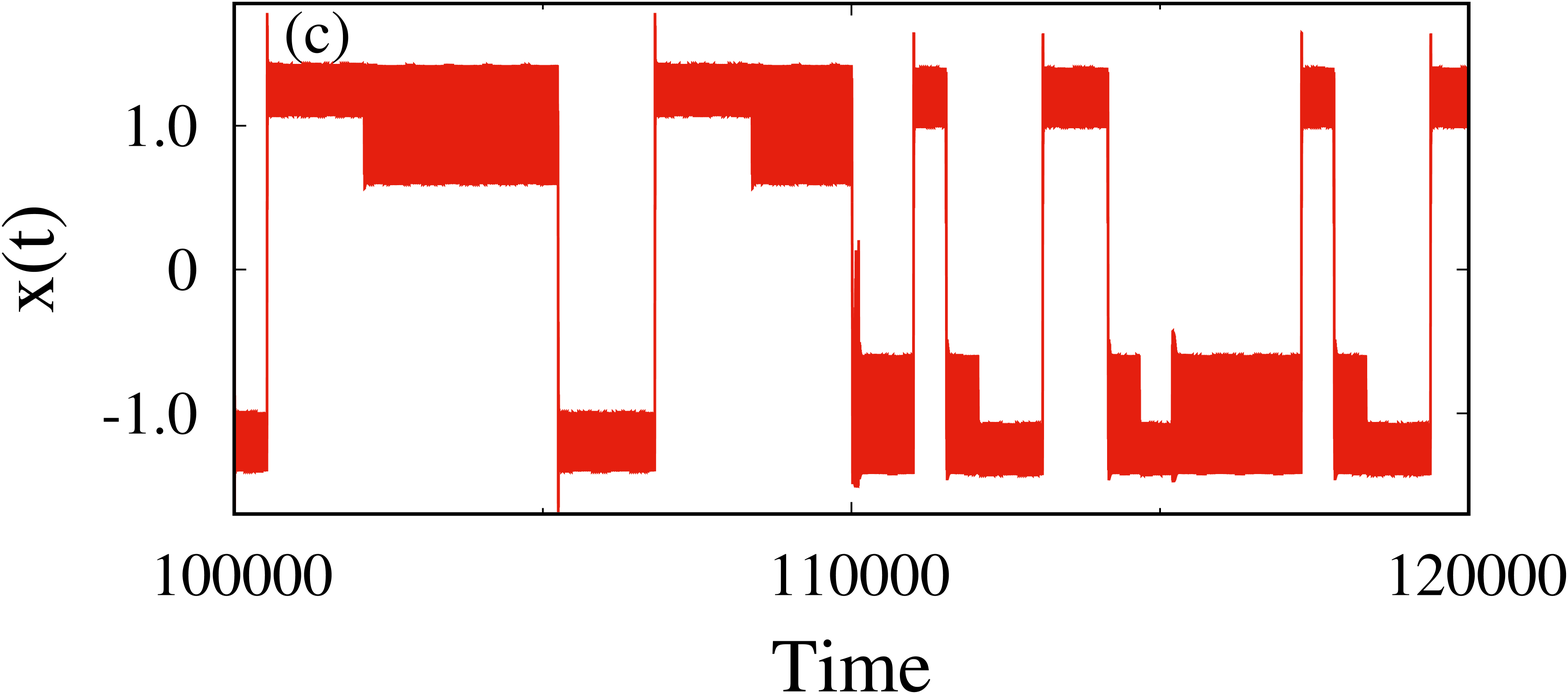}
	\caption{Panel (a) shows a combination of two input signals $I=I_{1}+I_{2}$. Input $I_{1}+I_{2}=+0.6$ when the logic input is `$1$' and $I_{1}+I_{2}=-0.6$ when the logic input is `$0$'. Panel (b) is the constant bias $ \varepsilon $, when it varies from $ 0.1 $ to $ -0.1 $. Panel (c) represents the corresponding dynamical response of the system x(t) under periodic forcing and noise with fixed parameters $F =0.44$ and $ D=0.00001 $ leading to OR logic gate (in the time range $ (1.0 * 10^5 - 1.1 * 10^5 ) $ and AND gate beyond $ 1.1 * 10^5 $ time units. }
	\label{fig9}
\end{figure}

\subsection{Effect of bias and implementation of different logic gates}

Now, we discuss how on changing the bias from positive to negative value in the optimum range, the dynamics of the system gets varied. For this purpose we study the effect of constant bias $ \varepsilon $ in \eqref{equ1}. From Fig.\ref{fig11}, it is obvious that for the input stream $ (1,1)$ the attractor is bounded in the $ x>0 $ state and it is in the $ x<0 $ state in the phase space for $ (0,0) $ state and for other input streams $ (1,0)/(0,1) $ it is again bounded in the $ x>0 $ region of the phase space when $ \varepsilon=0.1 $. Then the dynamical attractor will produce logical OR gate [see Fig.\ref{fig11}(a)]. On the other hand, when we change the bias to $ \varepsilon=-0.1 $ again, it is found that the response of the oscillator is in the $ x<0 $ region, so that the dynamical attractor will produce a logical AND gate [see Fig.\ref{fig11}(b)]. The logic output is `1' when $x(t)>0$ and `0' for $x(t)<0$. As a result, if we change the bias from $ \varepsilon=0.1 $ to $ \varepsilon=-0.1 $ the response of the system also changes from OR logic gate to AND logic gate as shown in Figs.\ref{fig9}. Fig.\ref{fig9}(a) shows the three level logic input $ I=I_{1}+I_{2} $. Fig.\ref{fig9}(b) represents the bias changing. In this diagram we observe that as the bias changes from positive to negative  optimum values the corresponding symmetry of the potential well  alternates, which leads to a change from OR logic gate to AND logic gate[Fig.\ref{fig9}(c)].

	\section{Conclusion}
	
	In the present paper, we have studied the existence of noise induced SNAs in a single periodically driven Duffing oscillator system. To test the validity of the robustness of such SNAs, we perturb the system by adding two logic signals which leads to the emulation of different logical behaviors. For appropriate input signals and noise, we realized different logical outputs in the above simple periodically driven nonlinear system. The present study is significant in the aspect that noise induced logical SNAs can be observed in \emph{periodically driven double-well Duffing oscillator with single periodic force}. Utilizing this feature, we have explicitly demonstrated the implementation of OR gate in the corresponding nonlinear system. On bias/threshold changing from positive to negative values the logic gate switches over from OR to AND logic gate.  Thus we see that noise assisted periodically driven system can exhibit logic behavior via logical SNA. Noise-induced SNAs, logical SNAs and standard SNAs are characterized by finite-time Lyapunov exponents, spectral characteristics and snapshot attractors to examine the strange and nonchaotic behaviors.
	 
	The logical stochastic resonance is realized only in an optimal range of noise strength. That is, it cannot be observed for low or high noise intervals. It is established that the structure of the attractor gets  smeared out by the large amplitude of noise. Thus attractors in random dynamical systems are purposeful only for small amplitude noise. Further, analog components in the electronic circuits generate noise in the microvolt range. Our study paves the way for implementation of logic elements only in the experimental regime of noise strength. Further in chaotic computing, a small amount of noise can make the computing to drastically change. Thus, our study also demonstrates that the existence of logical elements is possible in the SNA regime and that the phenomenon is robust in the presence of experimental noise. Thus noise induced SNA is a good candidate to realize reconfigurable, and flexible computer hardware. 
		
		\section*{AUTHOR’S CONTRIBUTIONS}		
		All authors contributed equally.		
		
		\section*{Acknowledgment} 
		M.S. sincerely thanks Council of Scientific \& Industrial Research, India for providing a fellowship under SRF Scheme No.08/711(0001)2K19-EMR-I.  A.V. is supported by the DST-SERB research project Grant No.EMR/2017/002813. M.L. acknowledges the financial support under the DST-SERB Distinguished Fellowship program under Grant No.SB/DF/04/2017.
		
		\section*{DATA AVAILABILITY}
		The data that support the findings of this study are available from the corresponding author upon reasonable request.

\end{document}